\documentclass[preprint,12pt]{elsarticle}

\usepackage{amssymb,amsmath,amsthm,amscd,amsfonts,mathrsfs}
\usepackage{caption}
\usepackage{subcaption}
\usepackage{tikz}
\usepackage{verbatim}
\usepackage{url}
\usepackage{fancyvrb}
\usepackage{listings}
\usepackage{inconsolata}
\usepackage{hyperref}

\def\mm {{\sc Mathematica}}
 \def\Singular{{\sc Singular}}
 \def\Azurite{{\sc Azurite}}
\def\la{\langle}
\def\ra{\rangle}
\def\nn{\nonumber}

\def\TF{{\mathbf T}}
\def\d{{\rm d}}
\lstdefinestyle{Common}
{
    basicstyle=\small \ttfamily\null,
    numbers=none,
    numbersep=1em,
    frame=single,
    framesep=\fboxsep,
    framerule=\fboxrule,
    xleftmargin=\dimexpr\fboxsep+\fboxrule,
    xrightmargin=\dimexpr\fboxsep+\fboxrule,
    breaklines=true,
    breakindent=0pt,
    tabsize=5,
    columns=flexible,
    showstringspaces=false,
    captionpos=b,
    abovecaptionskip=0.5\smallskipamount,   
}

\lstdefinestyle{Mathematica}
{
    style=Common,
    language={Mathematica},
    alsolanguage={[LaTeX]TeX},
    morekeywords=
    {
        Preparation,
        FeynmanGraph,
        MomentaSymmetry,
        PropagatorSymmetry,
        FindAllMIs,
        DiagramExtendedOutput,
        FetchCachedGraphInfo,
        DiagramCache, LoopMomenta, ExternalMomenta, Propagators,
        Kinematics, Numerics,
        DiagramAnalysis, FindAllMIs, Symmetries,
        mp2, IntegralRed, GlobalSymmetry, Symmetry, NumericMode,
        Characteristic, NumericD,HighestPower,WorkingPower, WatchingMode,SPExpand,
        TemporaryDirectory,SingularDirectory,ParallelMode
    },
}

\lstdefinestyle{MathPlain}
{
    style=Mathematica,
    keywordstyle=\textbf,
    captionpos=b,
}

\lstnewenvironment{MathPlain}[1][]
{\lstset{style=MathPlain,#1}}
{}

\newcommand{\lsm}[1]{\lstinline[style=MathPlain]{#1}}

\usepackage{amssymb}

\newcounter{bla}

\journal{Computer Physics Communications}

\begin{document}

\begin{frontmatter}

\title{\textsc{Azurite}: An algebraic geometry based package for finding bases of loop integrals}





\author[a,c]{Alessandro Georgoudis}\ead{Alessandro.Georgoudis@physics.uu.se}
\author[b,c]{Kasper J. Larsen}\ead{Kasper.Larsen@soton.ac.uk}
\author[c]{Yang Zhang\corref{author}}

\cortext[author] {Corresponding author. \textit{E-mail address:} \texttt{Yang.Zhang@phys.ethz.ch}}
\address[a]{Department of Physics and Astronomy, Uppsala University, SE-75108 Uppsala, Sweden}
\address[b]{School of Physics and Astronomy, University of Southampton, Highfield, Southampton, SO17 1BJ, United Kingdom}
\address[c]{ETH Z{\"u}rich, Wolfang-Pauli-Strasse 27, 8093 Z{\"u}rich, Switzerland}

\begin{abstract}
For any given Feynman graph, the set of integrals with all possible powers of the propagators spans a vector space of finite dimension. We introduce the package {\sc Azurite} ({\bf A ZUR}ich-bred method for finding master {\bf I}n{\bf TE}grals), which efficiently finds a basis of this vector space. It constructs the needed integration-by-parts (IBP) identities on  a set of generalized-unitarity cuts. It is based on syzygy computations  and analyses of the symmetries of the involved Feynman diagrams and is powered by the computer algebra systems {\sc Singular} and {\sc Mathematica}. It can moreover analytically calculate the part of the IBP identities that is supported on the cuts.
\end{abstract}


\end{frontmatter}



\noindent {\bf PROGRAM SUMMARY}

\begin{small}
\noindent
{\em Program Title:}  \textsc{Azurite}                                            \\
{\em Licensing provisions:} GNU General Public License (GPL)                      \\
{\em Programming language:}  Wolfram Mathematica version 10.0 or higher           \\
{\em Supplementary material:} A manual in the form of a Mathematica notebook      \\
{\em Nature of problem:} Determination of a basis of the space of loop integrals
                  spanned by a given Feynman diagram and all of its subdiagrams   \\
{\em Solution method:} Mathematica implementation
\end{small}

\section{Introduction}

Precision calculations of the cross sections of Standard Model
processes at the Large Hadron Collider (LHC) are crucial to gain a
quantitative understanding of the background and in turn improve the
ability to extract signals of new physics. This typically requires
computations at next-to-next-to leading order (NNLO) in fixed-order
perturbation theory, in order to match the experimental precision and
the parton distribution function uncertainties. Calculations at this
order are challenging because of the large number of contributing
Feynman diagrams, which involve loop integrals with high powers of
loop momenta in the numerator of the integrand.

A key tool in these calculations are integration-by-parts (IBP)
identities \cite{Tkachov:1981wb, Chetyrkin:1981qh}. These are relations that arise from the vanishing
integration of total derivatives. Schematically, they take the form,
\begin{equation}
  \label{IBP}
  \int \prod_{j=1}^L \bigg (\frac{\d^D l_j}{i \pi^{D/2}}\bigg)
  \sum_{i=1}^L \frac{\partial }{\partial l^\mu_i}
  \frac{v_i^\mu}{D_1^{a_1} \cdots D_k^{a_k}}=0\,,
\end{equation}
where the vectors $v_i^\mu$ are polynomials in the internal and
external momenta, the $D_k$ denote inverse propagators, and $a_i \geq 1$ are
integers. In practice, the IBP identities generate a large set of
linear relations between loop integrals, and allow a significant
fraction of them to be expressed in terms of a finite linear
basis. (The fact that the basis of integrals is always finite was
proven in ref.~\cite{Smirnov:2010hn}.) The latter step of solving the linear systems
arising from eq.~\eqref{IBP} may be carried out by Gauss-Jordan elimination in
the form of the Laporta algorithm \cite{Laporta:2001dd, Laporta:2000dc}, leading in general to
relations involving integrals with squared propagators. There are
several implementations of automated IBP reduction publically
available: AIR \cite{Anastasiou:2004vj}, FIRE \cite{Smirnov:2008iw,Smirnov:2014hma}, Reduze
\cite{Studerus:2009ye, vonManteuffel:2012yz}, LiteRed \cite{Lee:2012cn}, along
with private implementations. Finite field techniques can be used to speed up the computation \cite{vonManteuffel:2014ixa,vonManteuffel:2015gxa,vonManteuffel:2016xki, Peraro:2016wsq}.

A formalism for deriving IBP reductions that do not involve integrals
with squared propagators was developed in ref.~\cite{Gluza:2010ws},
based on syzygy computations. As observed in ref.~\cite{Schabinger:2011dz},
the syzygies can be computed with linear algebra methods.

In addition to reducing the contributing Feynman diagrams to
a small set of basis integrals, the IBP reductions provide
a way to compute these integrals themselves through differential
equations \cite{Kotikov:1990kg,Kotikov:1991pm,Bern:1993kr,Remiddi:1997ny,Gehrmann:1999as,Ablinger:2015tua}.
Letting $x_m$ denote a kinematical variable, $\epsilon = \frac{4-D}{2}$ the dimensional regulator,
and $\mathcal I({\mathbf x},\epsilon) =\{ I_1({\mathbf x},
\epsilon),\ldots,I_N({\mathbf x}, \epsilon)\}$ the basis of integrals, the
result of differentiating any basis integral wrt. $x_m$ can again be
written as a linear combination of the basis integrals by using, in
practice, the IBP reductions. As a result, one has a linear system
of differential equations,
\begin{equation}
  \label{DE}
  \frac{\partial }{\partial x_m} \mathcal I({\mathbf x},\epsilon) =A_m({\mathbf
    x},\epsilon)  \mathcal I({\mathbf x},\epsilon) \,,
\end{equation}
which, supplied with appropriate boundary conditions, can be solved to
yield expressions for the basis integrals. This has proven to be a
powerful tool for computing two- and higher-loop integrals. As observed in
ref.~\cite{Henn:2013pwa}, in many cases of interest, with an appropriate
choice of integral basis, the coefficient matrix $A_m$ in eq.~\eqref{DE}
becomes proportional to $\epsilon$. As a result, the basis integrals
are manifestly expressed as iterated integrals.
Refs.~\cite{Lee:2014ioa, Meyer:2016slj} provide algorithms for finding a transformation
to a canonical basis, which applies provided that a rational transformation
exists.\footnote{For some cases, the leading singularities are elliptic. Using
complete elliptic integrals, these differential equations can be solved
as iterated integrals with elliptic kernels
\cite{Remiddi:2016gno,Bonciani:2016qxi,Primo:2016ebd}.}

In many realistic multi-scale problems, such as $2 \to n$ scattering
amplitudes with $n \geq 2$, the step of generating IBP reductions with
existing algorithms is the most challenging part of the
calculation. It is therefore of interest to explore other methods for
generating these reductions.

In ref. \cite{Larsen:2015ped} a subset of the present authors showed
how IBP reductions that involve no squared propagators can be obtained
efficiently on specific (algorithmically determined) sets of
generalized-unitarity cuts. A similar approach was introduced by
Harald Ita in ref. \cite{Ita:2015tya} where IBP relations are also
studied in connection with cuts, and the underlying geometric interpretation is clarified.

In this paper we introduce the \Singular\ \cite{DGPS}/\mm\ package \Azurite\ ({\bf A ZUR}ich-bred method for finding master
{\bf I}n{\bf TE}grals) which determines a basis for the space of integrals
spanned by a given $L$-loop diagram and all of its subdiagrams
(obtained by shrinking propagators). \Azurite\ can also be
used to analytically generate IBP identities evaluated on maximal cuts.

In practice, the current version of this package can determine a basis of integrals for a
two-loop diagram and all of its subdiagrams (no matter whether massless or
massive, planar or non-planar) in seconds. It can also determine master
integrals for a three-loop diagram and all of its subdiagrams in minutes.

Related work has appeared in ref.~\cite{Lee:2013hzt} where the number of basis integrals is determined from the critical points of the polynomials that enter the parametric representation, or equivalently the Baikov representation, of the integral. This method has moreover been implemented in the Mathematica package Mint.

\section{Algorithm}
The algorithm of \Azurite\ may be summarized as follows: given an input diagram, the code traces over all subdiagrams and
\begin{enumerate}
\item automatically
determines the automorphism group of the involved Feynman diagrams by graph theory algorithms,
\item detects and discards scaleless integrals (for example, diagrams
  with massless tadpoles),
\item determines the linear relations between integrals evaluated on maximal cuts for each subdiagram, using
  the methods of ref.~\cite{Larsen:2015ped} of constructing the IBP
  identities on $D$-dimensional generalized-unitarity cuts and solving
  syzygy equations. The on-shell version of the IBP identities,
  which have been constructed so as to contain no integrals with
  higher-power propagators, are generated numerically
  via finite field computations in \Singular.
\end{enumerate}
After these steps, \Azurite\ chooses a basis of integrals according
to the following conventions: it removes all {\it edge-reducible}
integrals from the candidate list of master integrals. (An edge-reducible
integral is an integral which can be expressed as a linear combination
of integrals from its subdiagrams.) For the remaining integrals,
\Azurite\ considers IBP relations between integrals with different
numerators, and finds a linear basis of integrals which contains the
lowest possible numerator degrees. Only IBP identities evaluated on cuts are needed for determining the basis of integrals.

In the following we will explain the above steps in greater detail. To this end, we first introduce notation and some parametrizations of the integrals. We consider a general $L$-loop Feynman diagram with $n$ external lines, $k$
propagators, and all of its subdiagrams. The associated Feynman
integrals are,
\begin{equation}
  \label{Feynman}
  I[a_1,\ldots, a_k;N]\equiv \int \prod_{j=1}^L \bigg (\frac{\d^D l_j}{i \pi^{D/2}}\bigg)
  \frac{N(l_1,\ldots, l_L)}{D_1^{a_1} \cdots D_k^{a_k}} \, .
\end{equation}
Let $k_1,\ldots, k_n$ be
the external momenta, and $l_1,\ldots, l_L$ be the loop momenta. Following
ref.~\cite{Gluza:2010ws}, we restrict attention to IBP identities that do
not involve integrals with higher-power propagators. Moreover, we will ultimately
choose bases which do not contain such integrals, but rather contain integrals
with numerator insertions. Therefore we require for the indices that
$a_i \in \{0,1\}, i=1,\ldots, k$. To simplify the notation, we denote
\begin{equation}
  \label{eq:2}
  \la s_1 \ldots s_m\ra[N] \equiv \int \prod_{j=1}^L \bigg (\frac{\d^D l_j}{i \pi^{D/2}}\bigg)
  \frac{N(l_1,\ldots, l_L)}{D_{s_1} \cdots D_{s_m}} \,,
\end{equation}
where $1\leq s_1 < s_2 <\cdots <s_m\leq k$ are the indices for
existing propagators. We moreover use $ \la s_1 \ldots s_m\ra$ to denote
the topology of the corresponding subdiagram.

The inverse propagators take the generic form,
\begin{equation}
  \label{invProp}
  D_i= \Big(\sum_{j=1}^L \alpha_{ij} l_j +\sum_{h=1}^n \beta_{ih} k_h \Big)^2 -
  m_i^2 \hspace{1mm} \equiv \hspace{1mm} v_i^2-m_i^2 \,,
\end{equation}
where the $\alpha_{ij}$ and $\beta_{ih}$ are $\pm 1$.
$v_i$ denotes the momentum of the corresponding line.

We use dimensional regularization and work in the four-dimensional
helicity scheme, taking the external momenta to be strictly four-dimensional.
Accordingly, we decompose
the loop momenta into four- and $(D-4)$-dimensional parts,
$l_i = \overline{l}_i + l_i^\perp$. As explained in section 2 of ref.~\cite{Larsen:2015ped}, for $n\leq 4$, the external
momenta span a vector space of dimension less than four, and
the components of the loop momenta along the orthogonal directions
can be integrated out directly. After having done so, there are
\begin{equation}
  \label{SP_counting}
  n_\text{SP}=\phi(n) L +\frac{L(L+1)}{2} \,,
\end{equation}
independent scalar products involving the loop momenta, where
\begin{equation}
  \label{eq:7}
  \phi(n)\equiv \left\{
    \begin{array}{cc}
      4   & \hspace{5mm} n\geq 5 \,,\\
      n-1 & \hspace{5mm} n\leq 4 \,.
    \end{array}
\right.
\end{equation}
An application of the Ossola-Papadopoulos-Pittau (OPP) reduction method
\cite{Ossola:2006us,Ossola:2007ax,Ellis:2011cr,Ellis:2007br,Mastrolia:2011pr,Badger:2012dp}, or
integrand reduction via polynomial division wrt. Gr{\"o}bner bases
\cite{Zhang:2012ce,Mastrolia:2012an}, shows that if the number of distinct
propagators is greater than the number of independent scalar products;
i.e., $k>n_\text{SP}$, then
the diagram is reducible at the integrand level. Hence we can assume
without loss of generality that $k\leq n_\text{SP}$.

An important tool used in \Azurite\ is the Baikov representation \cite{Baikov:1996rk}
of an integral,
\begin{equation}
  \label{Baikov}
 \la 1 2 \ldots k \ra[N] \hspace{1mm} \propto \hspace{1mm} \int \d z_1 \cdots \d z_{n_\text{SP}} F(z_1,\ldots, z_{n_\text{SP}})^{\frac{D-h}{2}}
  \frac{N(z_1,\ldots, z_{n_\text{SP}})}{z_1 \cdots z_m} \,,
\end{equation}
where the $z_1, \ldots, z_m$ denote the inverse propagators
$D_{s_1},\ldots, D_{s_m}$. $z_{m+1}, \ldots, z_{n_\text{SP}'}$ denote
irreducible scalar products (ISPs, i.e., terms appearing in the numerator
which cannot be written as linear combinations of inverse propagators). The quantity $F \equiv \det_{i,j}
\mu_{ij}$, $i,j=1,\ldots,L$ appearing
in the measure factor is occasionally referred to as the Baikov polynomial,
whereas the exponent is defined as $h\equiv L+\phi(n)$. Here
\begin{equation}
\mu_{ij}\equiv -l_i^\perp \cdot l_j^\perp \,,
\end{equation}
where $l_i^\perp$ is the $(-2\epsilon)$-dimensional component of $l_i$.

This representation is particularly suitable
for generating IBP identities on generalized-unitarity cuts,
and was used in refs.~\cite{Ita:2015tya,Larsen:2015ped}.
\Azurite\ computes $F$ through an appropriate change
of variables of the loop momenta. It first parametrizes the loop
momenta via van Neerven-Vermaseren coordinates
\cite{vanNeerven:1983vr}, then separates the $\mu_{ij}$ and finally obtains the Baikov representation. 
The overall prefactor and the region of integration in eq.~\eqref{Baikov}
are irrelevant for deriving IBP identities, and hence we neglect these. (The expressions for the overall pre-factor of the Baikov representation can be found in ref. \cite{Lee:2010wea}.)

\subsection{Associated graphs and their symmetries}\label{graph_sym}

Given the propagators in eq.~\eqref{invProp}, it is useful to obtain
the corresponding graph algorithmically---i.e., to determine the
vertices---for the purpose of finding the discrete symmetries.
This can be achieved by a {\it backtracking} algorithm.
Define the set of flows of momenta on external and internal lines,
\begin{equation}
  \label{eq:3}
  M=\{k_1,\ldots, k_n, v_1,\ldots, v_k, -v_1, \ldots, -v_k\} \,.
\end{equation}
Search through the subsets of $M$ until finding a subset $V_1$ containing at least
three entries and satisfying momentum conservation, $\sum_{p\in V_1} p=0$. $V_1$ is the candidate for the
first vertex. Now redefine $M:=M-V_1$ and search through the subsets of $M$ to find
$V_2$ analogously. Iterate this process. If, at some step, no $V_i$ can be
found, then backtrack and redefine $M:=M\cup V_{i-1}$ and proceed to find a
new candidate $V_{i-1}'$ for the previous vertex. When $n_V$ vertices
have been found, and the resulting graph is connected and has $L$ loops, the
algorithm terminates. Here,
\begin{equation}
  \label{eq:4}
  n_V=k-L+1 \,,
\end{equation}
denotes the number of vertices (cf. section II.3 of ref. \cite{MR1633290}).

\begin{figure}
    \centering
    \begin{subfigure}[b]{0.4\textwidth}
        \includegraphics[width=\textwidth]{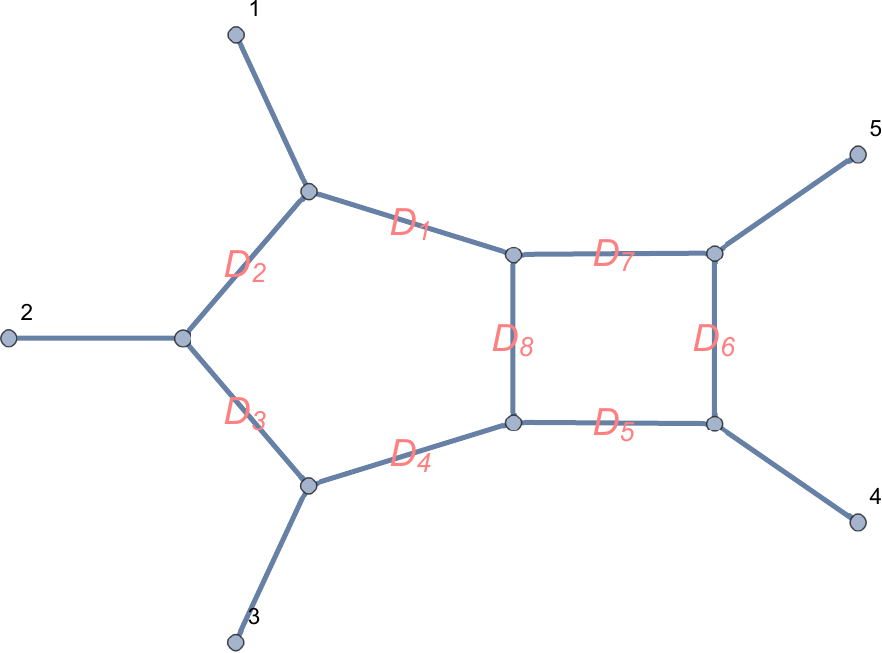}
        \caption{The pentagon-box diagram $\la 12345678\ra$.}
        \label{pentagon-box}
    \end{subfigure}
    \begin{subfigure}[b]{0.4\textwidth}
        \includegraphics[width=\textwidth]{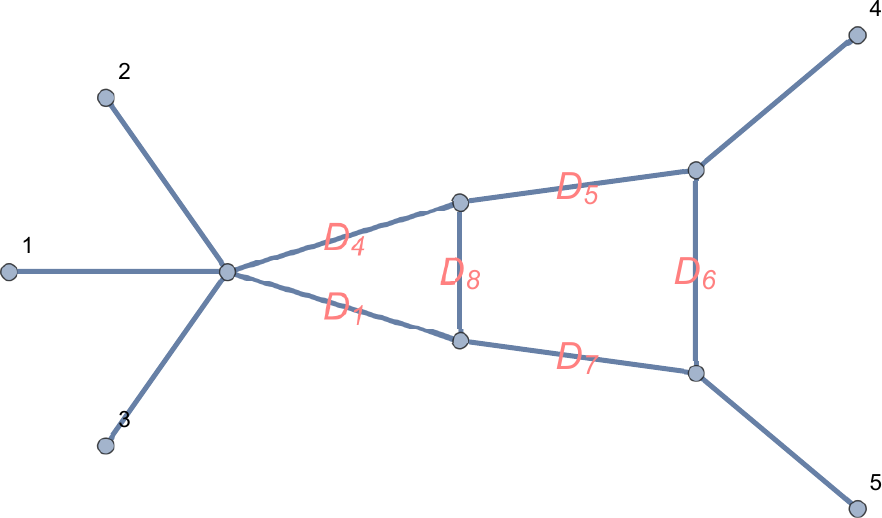}
        \caption{The $\la 145678\ra$ subdiagram of the pentagon-box diagram.}
        \label{triangle-box}
    \end{subfigure}
\caption{Pentagon-box diagram and one of its subdiagrams.}\label{fig:pentabox_family}
\end{figure}

As an example let us consider $L=2$ and the following eight inverse propagators,
\begin{equation}
\begin{alignedat}{4}
  & D_1 = l_1^2\,,            \hspace*{0.27cm}  && D_2 = (l_1 - k_1)^2\,,  \hspace*{0.27cm}  && D_3 = (l_1 - K_{12})^2\,,  \hspace*{0.27cm}  && D_4 = (l_1 - K_{123})^2 \,, \\
  & D_5 = (l_2 + K_{123})^2\,,  \hspace*{0.27cm}   && D_6 = (l_2 -k_5)^2\,,   \hspace*{0.27cm}  && D_7 = l_2^2\,,             \hspace*{0.27cm}  && D_8 = (l_1 + l_2)^2 \,.
\label{pentabox_invProp}
\end{alignedat}
\end{equation}
where $K_{i_1\cdots i_s}\equiv k_{i_1}+\cdots + k_{i_s}$.
The backtracking method finds the vertices,
\begin{equation}
\begin{alignedat}{2}
  & V_1=\{k_1,-l_1,l_1-k_1\}\,,              \quad  && V_2=\{k_2,l_1 - K_{12},-l_1+k_1\}\,,\\
  & V_3=\{k_3,l_1-K_{123},-l_1+K_{12}\}\,,   \quad  && V_4=\{k_4,l_2+K_{123},-l_2+k_5\}\,, \\
  & V_5=\{k_5,-l_2,l_2-k_5\}\,,              \quad  && V_6=\{-(l_1+l_2),l_1,l_2\}\,, \\
  & V_7=\{-l_1+K_{123}, -(l_2+K_{123}),l_1+l_2\} \,. \quad && \phantom{V_8={}}
\label{eq:5}
\end{alignedat}
\end{equation}
From this information it is straightforward to construct
the adjacency matrix of the graph. The graph is found to be
the pentagon-box diagram illustrated in fig.~\ref{pentagon-box}.

Once the graph for $\la 12\ldots k\ra$ has been found, all of its
subdiagrams can be obtained by pinching subsets of its propagators.
Taking a graph theoretical viewpoint, we obtain the various subdiagrams
by appropriately truncating the adjacency matrix of the original graph.\footnote{For example, if the edge $e$ connecting two vertices $v_1$ and $v_2$ is pinched, then we merge the two columns (and also the two rows) in the adjacency matrix that  correspond to $v_1$ and $v_2$. } For example,
$\la 12345678\ra$ in fig.~\ref{pentagon-box} corresponds to
a pentagon-box diagram, and $\la 145678\ra$ in fig.~\ref{triangle-box} corresponds to to a
triangle-box subdiagram.

Given a subdiagram $\la s_1s_2\ldots s_m\ra$, after having obtained its graph,
we proceed to find its {\it automorphism group} $G'$ via a
graph theory based algorithm. $G'$ acts on the propagators. Let $G$ denote
the subgroup of $G'$ which preserves all external Lorentz invariants.
$G$ is the physical symmetry group of this diagram.
$G$ actually classifies all subdiagrams of  $\la s_1s_2\ldots s_m\ra$ into equivalence
classes. (If two subdiagrams $g_1$ and $g_2$ are equivalent, then any
integral with the topology $g_1$ must equal
an integral with the topology $g_2$ with the appropriate numerator insertion.)

Furthermore, $G$ acts on momenta as affine
transformations (linear and shift transformations). We explicitly find
these transformations by linear algebra. This enables us to determine
the action of $G$ on irreducible scalar products appearing in the
numerator.

For instance,  diagram $\la 145678\ra$ associated with the inverse propagators in eq.~\eqref{pentabox_invProp}
has the symmetry group $G=Z_2$, whose non-trivial element is (cf.~fig.~\ref{triangle-box}),
\begin{gather}
  \label{145678_G}
  D_1 \mapsto D_4\,, \quad D_4 \mapsto D_1 \,,\quad D_5 \mapsto D_7\,,\quad
  D_7 \mapsto D_5\,, \quad D_8 \mapsto D_8 \,.
\end{gather}
(Note that since both $k_4$ and $k_5$ are massless this symmetry preserves external Lorentz invariants,
and hence is physical.)
This implies symmetry relations for all of its subdiagrams. For
example, by this symmetry, the integral $\la 158\ra$ is equal
to $\la 478\ra$. Thus, we only need to consider $\la 158\ra$ during
the search for master integrals, and can neglect $\la 478\ra$.

In this example, the non-trivial element of $G$ given in eq.~\eqref{145678_G}
corresponds to the affine transformation,
\begin{gather}
  \label{145678_Affine}
  k_4 \mapsto k_5\,, \quad k_5 \mapsto k_4\,, \quad l_1 \mapsto -k_4-k_5-l_1\,, \quad l_2 \mapsto -k_4-k_5-l_2\,.
\end{gather}
From this, the action of $G$ on numerator polynomials can readily be found.

This backtracking graph-construction algorithm is implemented in
\Azurite, powered by \mm. The graph automorphism groups,
connectedness condition, and other graph information are computed
via Mathematica's embedded graph commands.\footnote{\mm\ 10.0.0 or later versions are required for the graph theory computations in \Azurite.} The affine transformations
such as those in eq.~\eqref{145678_Affine} are obtained by setting up an ansatz
of the action on the momenta of the internal lines,
\begin{equation}
  \label{eq:6}
  v_i \mapsto c_i v_{g(i)}\,,\quad \forall i\in\{s_1,\ldots, s_m\}
\end{equation}
for $g \in G$. To ensure that this is a permutation of the propagators,
all of the $c_i$ must be $\pm 1$. Using standard linear algebra techniques,
the values of the $c_i$ are readily solved for, and the affine transformation is
determined.

\subsection{Adaptive parameterization and further graph simplifications}\label{adp_red}

\begin{figure}
    \centering
    \begin{subfigure}[b]{0.4\textwidth}
        \includegraphics[width=\textwidth]{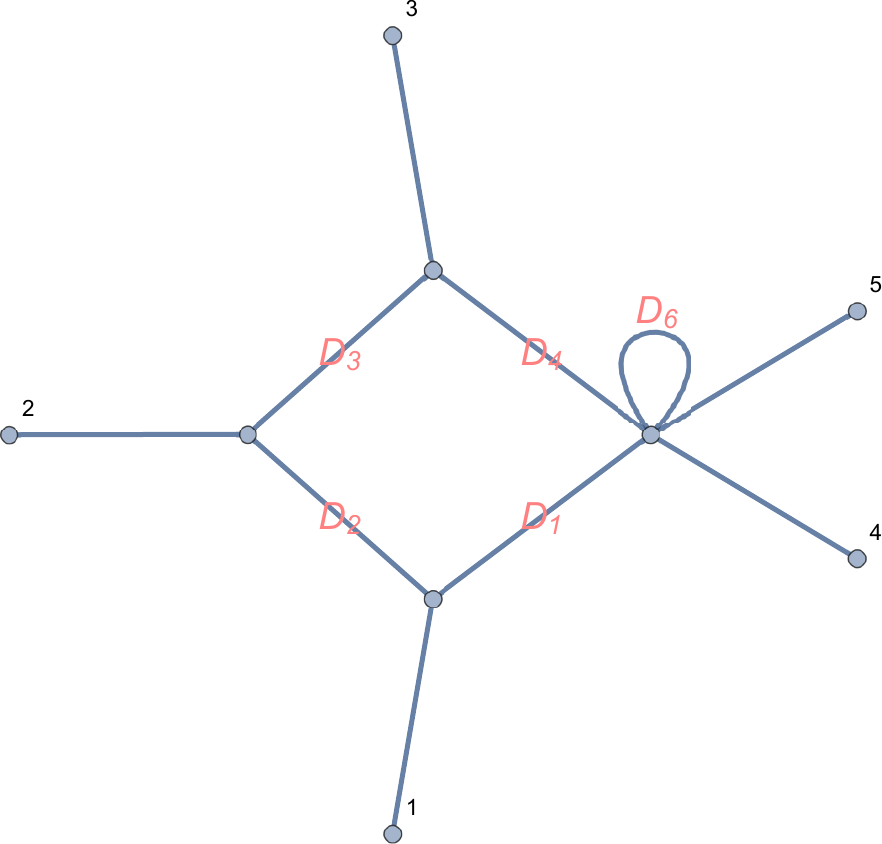}
        \caption{The subdiagram $\la 12346\ra$, which contains a massless tadpole.}
        \label{tadpole}
    \end{subfigure}
    \begin{subfigure}[b]{0.4\textwidth}
        \includegraphics[width=\textwidth]{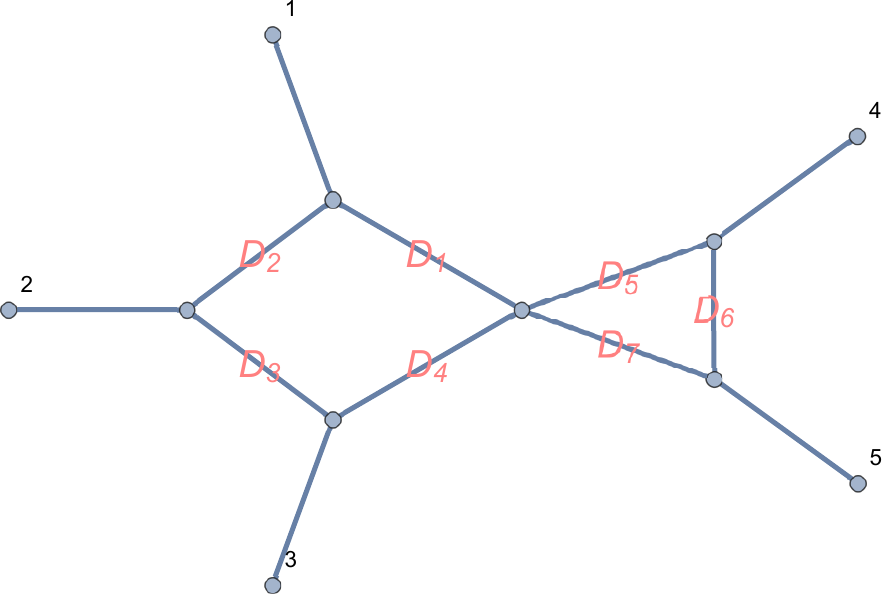}
        \caption{The subdiagram $\la 1234567\ra$, which is factorable.}
        \label{factorable}
    \end{subfigure}
\caption{Some diagrams which can be simplified in the adaptive parametrization.}\label{fig:pentabox_family}
\end{figure}

To optimize the search for master integrals we apply the following
simplifications during the study of the input diagram and all of its
subdiagrams. (Similar simplifications for subdiagrams are used in the adaptive integrand decomposition approach
of ref.~\cite{Mastrolia:2016dhn}.)
\begin{enumerate}
\item If a diagram has a loop which corresponds to a scaleless
  integral, then the diagram vanishes in dimensional regularization.
  For example, with the inverse propagators in eq.~\eqref{pentabox_invProp}, the diagram $\la
  12346\ra$ (illustrated in fig.~\ref{tadpole}) contains a massless tadpole and hence vanishes.
  \Azurite\ finds such loops by examining the fundamental
  cycles\footnote{See section II.3 of ref.
  \cite{MR1633290} for the definition of fundamental cycles.} of
  the graph.
  Moreover, the diagram $\la 1234 \ra$ corresponds to an integral
  without $l_2$ appearing in the denominator,
  so that the $l_2$ integral is scaleless and hence vanishes. Both of
  these diagrams are therefore discarded.

\item If for a diagram, two or more external lines attach to one
  vertex, we may combine these external lines into one external line
  with the sum of the individual momenta flowing on it. We let
  $n'$ denote the number of new external lines after this
  procedure, where clearly $n'<n$. As an example, for the inverse propagators in eq.~\eqref{pentabox_invProp}, the diagram $\la
  145678\ra$ (illustrated in fig.~\ref{factorable}) can be treated as a three-point diagram with the new external
  momenta $K_{123}$, $k_4$ and $k_5$.  It may occasionally be necessary to shift the loop momenta to
  ensure that only the new external momenta appear in the
  propagators. This is achieved in \Azurite\ by linear algebra methods.

We also define
  $n_\text{SP}'=\phi(n')L+L(L+1)/2$ as the number of new independent scalar
  products. For example, the diagram $\la
  145678\ra$ (illustrated in fig.~\ref{factorable}) has $n_\text{SP}'=2\times 2+3=7$.
This process decreases the number of scalar products and thereby
  significantly speeds up the IBP computations.

\item If a diagram consists of $n_\Gamma$ ($n_\Gamma>1$) loops that do not share common
  edges, we call the diagram {\it factorable} and treat the
  corresponding integral as a product
  of $n_\Gamma$ integrals. For example, with the inverse propagators given in eq.~\eqref{pentabox_invProp}, the diagram $\la
  1234567\ra$ is treated as the product of two one-loop diagrams. This
  is achieved in \Azurite\ by examining the fundamental cycles of the graph.
\end{enumerate}

For any subdiagram $\la s_1 \ldots s_m\ra$ encountered we apply these simplifications.
After doing so, the diagram has $n'$ external lines and $n_\text{SP}'$ independent scalar
products, and we can assume without loss of generality that the diagram
is non-factorizable and contains no scaleless integrals. By integrand reduction,
we furthermore have $m\leq n_\text{SP}'$. We then proceed to cast the corresponding
integrals in their Baikov representation \eqref{Baikov}
(with the number of independent scalar products
computed from the adaptive parametrization of the integral so that
$n_\mathrm{SP} \to n'_\mathrm{SP}$ and $h = L + \phi(n')$).

\subsection{IBP identities on maximal cuts and master integrals}
Given the input diagram $\la 12\ldots k\ra$, let us denote the set which consists of $\la 12\ldots k\ra$ and all of its
subdiagrams as $\mathcal S''$. Using symmetries we identify equivalent
diagrams within $\mathcal S''$ and obtain the subset $\mathcal S' \subset
\mathcal S''$ such that no two elements of $\mathcal S'$ are equivalent by a discrete symmetry.
$\mathcal S'$ consists of candidate topologies for master integrals.

Furthermore,
we discard diagrams in $\mathcal S'$ with scaleless loops, and simplify diagrams by rewriting them in their
adaptive representation if applicable, as described in the previous
subsection. The set of remaining diagrams is denoted by $\mathcal
S$. Then we
cast the integrals in $\mathcal{S}$ in their Baikov representation (cf. eq. \eqref{Baikov}).


There are many different ways of choosing a basis of integrals.
\Azurite\ chooses the basis as follows: it prefers
integrals with monomials in the numerator to integrals with higher-power propagators.
Moreover, it prefers a choice of basis whose integrals have as few propagators as possible.
This is to make the computation more efficient, as this
convention facilitates the use of IBP relations evaluated on their
maximal cuts. (Otherwise, we need complete IBP relations, without cuts applied, to find an
integral basis.)
Accordingly, as the code traces over the subdiagrams of the input
diagram, it removes {\it edge-reducible integrals}. These are integrals
which can be expressed as a linear combination of integrals that
correspond to strict subdiagrams. Evaluated on its maximal cut $D_{s_1} = \cdots = D_{s_m} = 0$,
an edge-reducible integral reads,
\begin{equation}
  \label{collapsible}
  \la s_1 \ldots s_m\ra[N] \hspace{1mm}=\hspace{1mm} 0 + \mbox{(strict subdiagrams)} \,,
\end{equation}
where the strict subdiagrams vanish on this cut.
$N$ is a monomial of irreducible scalar products.
Similarly, for the remaining integrals, we consider IBP identities without squared propagators
(cf.~ref.~\cite{Gluza:2010ws}) to find linear relations between integrals with different numerators. 
Again, evaluated on its maximal cut, a general IBP identity reads,
\begin{equation}
  \label{monomial_IBP}
  \sum_i c_i \la s_1 \ldots s_m\ra[N_i]= (\cdots )\,,
\end{equation}
where each $N_i$ is a monomial, and $(\cdots)$ denotes integrals that
correspond to strict subdiagrams. We moreover use the symmetry group
$G$ of $\la s_1 \ldots s_m\ra$ to find linear relations, taking a form
similar to that of eq.~\eqref{monomial_IBP}. To find the linear basis
of integrals, we introduce a monomial order $\succ$ for all monomials
in the irreducible scalar products.  After obtaining enough IBP and symmetry relations, we linearly reduce integrals according to $\succ$ via Gaussian elimination. In practice, a good choice of $\succ$ is either {\it degree reverse lexicographic} or {\it degree lexicographic} order,
as this ensures that the chosen basis contains monomials with as low
degree as possible. (In contrast, lexicographic monomial order may
lead to high-degree numerators.)

\Azurite\ traces through all diagrams in $\mathcal S$ and obtains the
complete list of master integrals. Because of the nature of sub-diagrams,
this computation can be finished in a parallelized way. Only IBP identities evaluated on their maximal cuts and symmetry relations are needed to find a basis.
Hence we focus our attention to obtaining eqs.~\eqref{collapsible} and \eqref{monomial_IBP},
i.e., IBP identities evaluated on their maximal cut. The representation
in eq.~\eqref{Baikov} can easily accommodate the maximal cut of any
subdiagram by taking the residue at $z_1= \cdots =z_m=0$, (adaptive
parametrization is used so that $z_1,\ldots, z_m$ denote
propagators and $z_{m+1},\ldots, z_{n_\text{SP}'}$ denote ISPs)
\begin{align}
  \label{Baikov_mc}
 \la s_1 \ldots s_m\ra[N] \Big|_\text{maximal cut} \hspace{2mm} &\propto \hspace{2mm}
 \int \d z_{m+1} \cdots \d z_{n_\text{SP}'} F(0,\ldots,0,z_{m+1},\ldots , z_{n_\text{SP}'})^{\frac{D-h}{2}}\nn\\
  &\hspace{20mm} \times N(0,\ldots,0,z_{m+1},\ldots , z_{n_\text{SP}'}) \,.
\end{align}
Again we neglect the overall prefactor and the region of integration. Define
$f(z_{m+1},\ldots , z_{n_\text{SP}'})\equiv
F(0,\ldots,0,z_{m+1},\ldots , z_{n_\text{SP}'})$.
Cf. refs. \cite{Ita:2015tya,Larsen:2015ped}, IBP identities evaluated on their maximal cut take the following form,\footnote{Note that the third term of the general form given in eq.~(11) of ref.~\cite{Larsen:2015ped}
is absent on the maximal cut where the number of cuts is equal to the number of propagators, $c=k$.}
\begin{align}
  0&=\int
 \d z_{m+1} \cdots \d z_{n_\text{SP}'} \sum_{i=m+1}^{n_\text{SP}'}
 \frac{\partial }{\partial z_i} \bigg( a_i(z_{m+1},\ldots , z_{n_\text{SP}'})
 f(z_{m+1},\ldots , z_{n_\text{SP}'})^\frac{D-h}{2}\bigg) \\
&=\int
 \d z_{m+1} \cdots \d z_{n_\text{SP}'}  \bigg(f^\frac{D-h}{2}\sum_{i=m+1}^{n_\text{SP}'}
 \frac{\partial  a_i}{\partial z_i}
 +\frac{D-h}{2}f^\frac{D-h-2}{2}\sum_{i=m+1}^{n_\text{SP}'} a_i
 \frac{\partial  f}{\partial z_i}\bigg) \,.
\label{IBP_Ansatz}
\end{align}
Here the $a_i$ (a priori) are arbitrary polynomials in the ISPs $z_{m+1},\ldots,
z_{n_\text{SP}'}$. The second term in eq.~\eqref{IBP_Ansatz} corresponds
to integrals in $D-2$ dimensions. To compensate this shift we
require,
\begin{gather}
  \label{syzygy}
  \sum_{i=m+1}^{n_\text{SP}'} a_i \frac{\partial f}{\partial z_i} +a f=0 \,,
\end{gather}
where $a$ is a polynomial in the ISPs. Equations of this kind are
known in algebraic geometry as \emph{syzygy equations}. Syzygy equations were also used for deriving IBP identities for integrals in Feynman parametrization \cite{Lee:2014tja}. The current version of \Azurite\
uses the command \pmb{syz} in \Singular\, to
find all generators of the solution set of eq.~\eqref{syzygy}.

Then the IBP identity evaluated on its maximal cut reads,
\begin{gather}
  \label{IBP_maximal_cut}
 0= \int
 \d z_{m+1} \cdots \d z_{n_\text{SP}'} f^\frac{D-h}{2} \bigg(\sum_{i=m+1}^{n_\text{SP}'}
 \frac{\partial a_i}{\partial z_i}
 -\frac{D-h}{2}a\bigg) \,,
\end{gather}
or equivalently,
\begin{equation}
  \label{eq:9}
  \la s_1 \ldots s_m\ra \bigg[\sum_{i=m+1}^{n_\text{SP}'}
 \frac{\partial  a_i}{\partial z_i}
 -\frac{D-h}{2}a\bigg] = (\cdots) \,,
\end{equation}
where $(\cdots)$ denotes integrals that correspond to strict subdiagrams.

In practice, given the generators of the syzygy module,
\begin{gather}
  \label{eq:10}
  \mathbf g^{(j)}=(a_{m+1}^{(j)},\ldots a_{n_\text{SP}'}^{(j)},a^{(j)}),
\end{gather}
we need to consider the syzygy $(a_{m+1},\ldots a_{n_\text{SP}'},a)=P
\mathbf g^{(j)}$ for the IBP formula \eqref{IBP_maximal_cut}. Here $P$ is an arbitrary
polynomial in the ISPs, with the degree up to a fixed integer.

\Azurite\ generates IBP identities evaluated on their maximal cut by
the use of eq.~\eqref{IBP_maximal_cut} allowing IBP identities up to a maximum degree. For the Gaussian elimination step, it lists the coefficients
of monomials in each IBP identity in a monomial order of ISPs (degree
reverse lexicographic by default). In this way a matrix of IBP
coefficients is obtained. Then by Gaussian elimination of this matrix, independent IBP identities are
identified. The master integrals correspond to the non-pivot columns
of the reduced matrix. The current version of \Azurite\ uses \pmb{slimgb} in \Singular, which
applies fast sparse linear algebra algorithms to carry out Gaussian
elimination. The integral basis search can be parallelized for the
sub-diagrams in $\mathcal S$, via the command \pmb{ParallelTable} in \mm.

For the purpose of finding a basis of integrals, numerical values for
the external kinematic invariants and spacetime dimension suffice. Using in addition finite field techniques, this has the benefit of speeding up the computation of syzygies and the Gauss-Jordan elimination step.
In some cases analytic IBP identities evaluated on maximal cuts are useful,
for instance for the study of multi-loop maximal unitarity in integer spacetime dimensions
\cite{Kosower:2011ty,Johansson:2012sf,CaronHuot:2012ab,
Johansson:2012zv,Johansson:2013sda,Sogaard:2013fpa}.
In this case analytic kinematics and spacetime dimension
$D$ would be used by \Azurite\ for generating analytic IBP identities on the maximal cut.

\subsection{Geometric interpretation of syzygy equation}

In this subsection we digress from the mainstream of the text
to discuss a geometric interpretation of the
constraint \eqref{syzygy}. The geometric picture of syzygies evaluated on
unitarity cuts was first discussed in ref.~\cite{Ita:2015tya}. Here we
reformulate the geometric interpretation in {\it tangent algebra} language.

The basic observation is that
the polynomial-valued vector field
\begin{gather}
  \label{eq:8}
  \sum_{i=m+1}^{n_\text{SP}'}  a_i \frac{\partial}{\partial z_i} \,,
\end{gather}
is tangent to the hypersurface defined by $f(z_{m+1},\ldots ,
z_{n_\text{SP}'})=0$ \cite{Ita:2015tya}. The solution set of
eq.~\eqref{syzygy} is the module of syzygies,
\begin{gather}
  \label{eq:1}
  \mbox{syz} \bigg(\frac{\partial f}{\partial z_{m+1}},\ldots, \frac{\partial
      f}{\partial z_{n_\text{SP}'}},f \bigg) \,.
\end{gather}
The $(a_{m+1},\ldots, a_{n_\text{SP}'})$ from this syzygy module form the
module of the {\it tangent algebra} $\TF_f$ \cite{Hauser1993}, i.e.,
the set of all polynomial-valued tangent vector fields for the
hypersurface $f=0$. $\TF_f$ is a Lie
algebra and infinite-dimensional in general.

The structure of $\TF_f$ depends on the geometric properties of the
hypersurface $f=0$. For example, when the hypersurface is {\it
non-singular}, i.e., the {\it singular ideal} $I_s$ satisfies
\begin{equation}
  \label{non-singular-hypersurface}
  I_s\equiv \bigg\la \frac{\partial f}{\partial z_{m+1}},\ldots, \frac{\partial
      f}{\partial z_{n_\text{SP}'}},f \bigg\ra =\la 1 \ra\,,
\end{equation}
then the solution of eq.~\eqref{syzygy} is
generated by {\it principal syzygies} (trivial syzygy relations) \cite{opac-b1094391}.
This can be proven by multiplying any syzygy relation by ``$1$'', and
replacing ``$1$'' by the generators of the singular ideal in eq. \eqref{non-singular-hypersurface}.
 In this case, no computation is
needed for obtaining the generators of $\TF_f$.

If the hypersurface $f=0$ is singular, then locally around
a singular point, $\TF_f$ is generated by principal syzygies and
{\it weighted Euler vectors} \cite{Hauser1993}. Moreover,
 cf. Schreyer's theorem \cite{opac-b1094391},
the generators of the solutions of eq.~\eqref{syzygy} can
be found algebraically via S-polynomial
computations.

\section{Examples and performance}

In this section we present some non-trivial results obtained from {\sc Azurite}
along with some benchmarks of its performance.
An introduction to the functions and their usage can be found in \ref{usage_azu}.
In all of the following cases, the full numerical approach is
used.\footnote{\label{pcsteup}The computations were carried out on a
  i7-6700, 32GB DDR4 RAM machine using \Singular ~v4.0.3, with
  parallel computations.} This setup is the most computationally favourable for \Singular.

\begin{figure}[h]
    \centering
    \includegraphics[width=\textwidth]{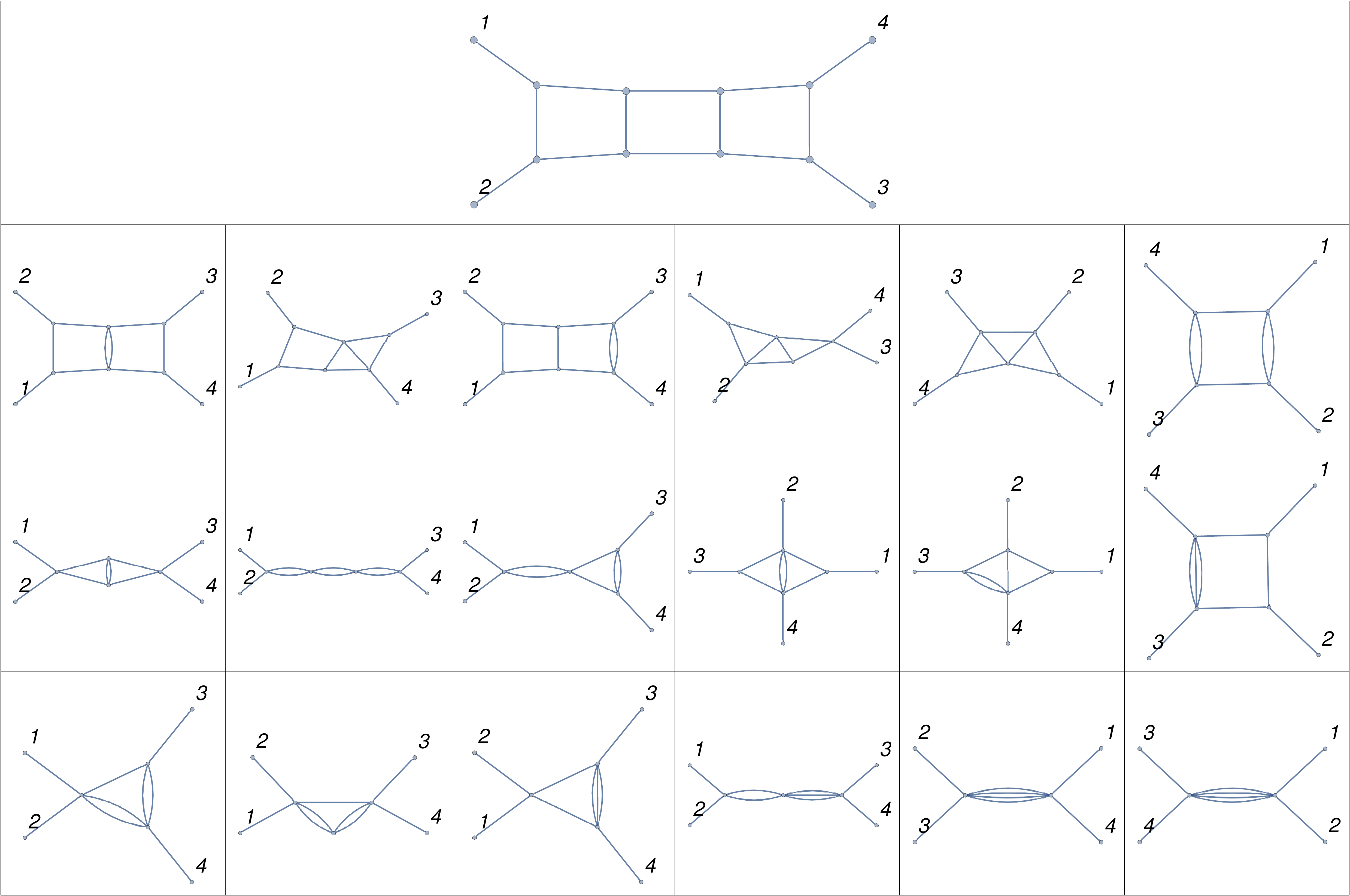}
    \caption{Irreducible topologies for the massless triple-box diagram.}\label{tb_fg}
\end{figure}

As an example, let us consider the triple-box diagram with $k=10$ propagators
illustrated in the top row of fig.~\ref{tb_fg}. First we present the initialization
of {\sc Azurite} for this diagram, shown in the sample code \ref{tb_def}.

The list of loop momenta is declared
in \lsm{LoopMomenta}. A list of linearly independent momenta is declared in
\lsm{ExternalMomenta}. The list \lsm{Propagators} consists of the propagators
of the diagram, augmented by a list of the independent ISPs. They are found by enumerating all the possible scalar products involving the loop momenta, and by finding a maximum-rank subset. In the case at hand there are, cf.~eq.~\eqref{SP_counting}, $n_\text{SP} - k = 15 - 10 = 5$ independent ISPs. These are the last five elements of \lsm{Propagators} below. In \lsm{Kinematics},
Lorentz invariants formed of external momenta are expressed in terms of the Mandelstam invariants.
In \lsm{Numerics}, numerical values are given for the kinematical invariants. These must
be chosen randomly, so as to avoid poles in the intermediate reduction steps.


\begin{MathPlain}[float,floatplacement=H, caption=Initialization for a massless triple-box diagram.,label=tb_def]
LoopMomenta = {l1, l2, l3};
ExternalMomenta = {k1, k2, k4};
Propagators = {l1^2, (l1 - k1)^2, (l1 - k1 - k2)^2,   l3^2, (-l3 - k1 - k2)^2, (l1 + l3)^2, (l2 - l3)^2,  l2^2, (l2 - k4)^2, (l2 + k1 + k2)^2, (l1 + k4)^2, (l2 + k1)^2, (l3 + k1)^2, (l3 + k4)^2, (l1 + l2)^2};
Kinematics = {k1^2 -> 0, k2^2 -> 0, k4^2 -> 0, k1 k2 -> s/2,  k2 k4 -> (-s - t)/2, k1 k4 -> t/2};
Numerics = {s -> 1, t -> -6};
Symmetries = {};
Preparation[];
\end{MathPlain}

Having declared the diagram, we can now proceed to compute the master integrals
of the vector space spanned by this diagram and its subdiagrams. This is done
with the \lsm{FindAllMIs} function,
\begin{MathPlain}[float,floatplacement=H,caption=Example of use for \lsm{FindAllMIs}.,label=tripleboxMI]
MIs=FindAllMIs[{1,2,3,4,5,6,7,8,9,10},NumericMode -> True,NumericD -> 1119/37,Characteristic -> 9001,HighestPower -> 3,WorkingPower -> 3,Symmetry -> True]
\end{MathPlain}
where the first input entry of \lsm{FindAllMIs} (sample code
\ref{tripleboxMI}) is a list of labels of the propagators
of the diagram. \lsm{FindAllMIs} can be used with the parallel computation.
We refer to section~\ref{appendix_mi} for further details on the syntax.

The computation is performed by making use of adaptive parametrizations
(cf.~section~\ref{adp_red}) of all the subdiagrams encountered in the IBP relations
that are generated, and taking into account their discrete symmetries (cf.~section~\ref{graph_sym}).
With the options chosen above, the computation is moreover performed in a
finite field of characteristic $9001$ and with the numerical value of $\frac{1119}{37}$
for the space-time dimension, chosen such that there are no dimension
dependent poles in the reduction coefficients.

The total time elapsed for the complete reduction is, on our desktop
computer with parallel computation
, 68 seconds. The irreducible topologies that are chosen as a basis are shown in fig.~\ref{tb_fg}. Their respective graphs were drawn using the function \lsm{FeynmanGraph}.

\begin{figure}[h]
    \centering
    \includegraphics[width=\textwidth]{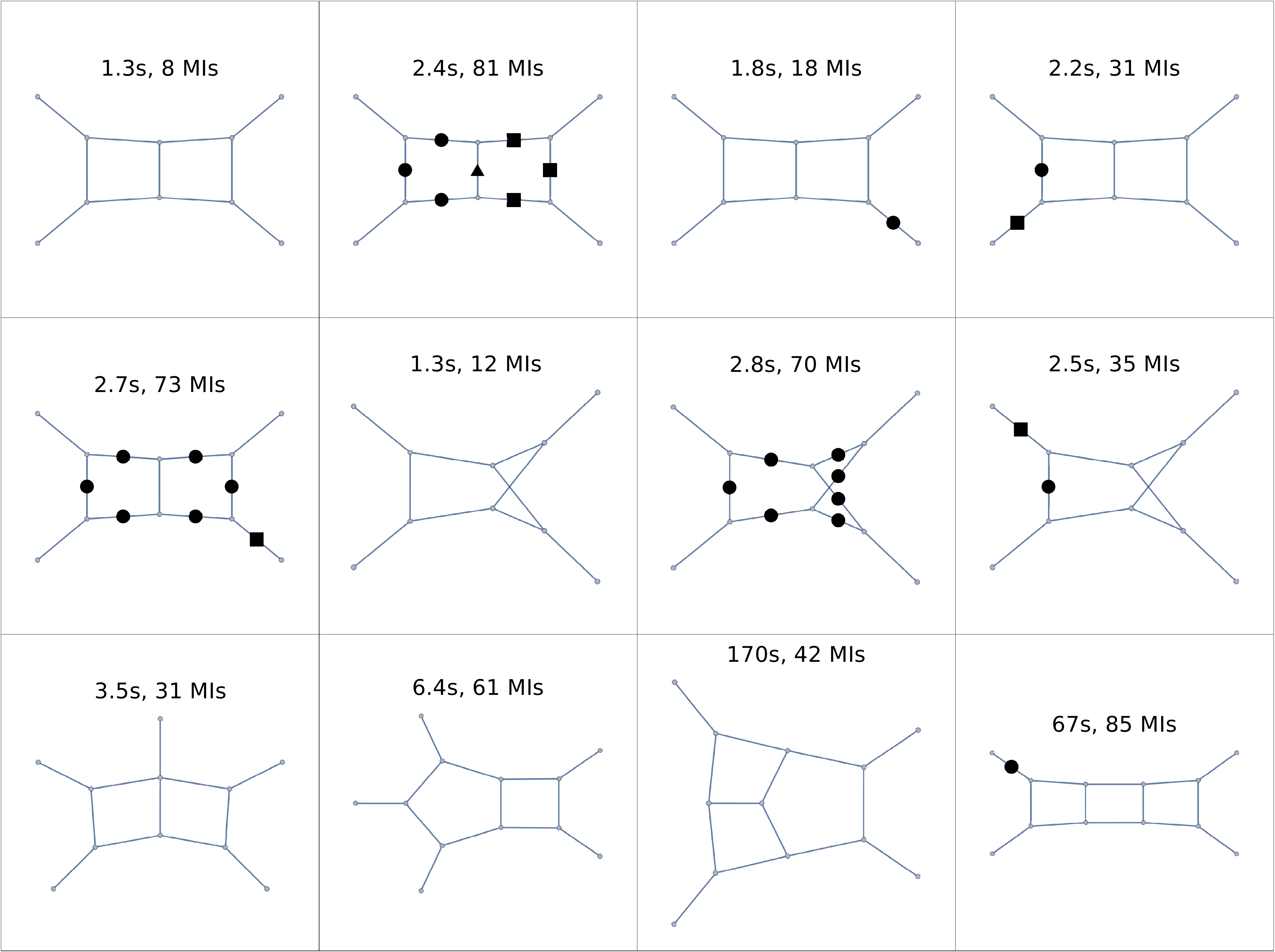}
    \caption{Computation time and number of master integrals for different topologies and mass configurations.}\label{ex_rd}
\end{figure}

Using the notation of eq.~\eqref{Baikov} we can rewrite the 26 basis elements as
\begin{align}\label{mi3b}
&\langle 123456789\hspace{0.8mm}10 \rangle [ z_{11} , z_{13} , 1 ] \quad \langle 1236789\hspace{0.8mm}10 \rangle [ z_4, 1 ] \quad \langle 1234679\hspace{0.8mm}10 \rangle [ z_5, 1 ] \nonumber \\
& \langle 12345679 \rangle [ z_8, 1 ] \quad \langle 125678\hspace{0.8mm}10 \rangle [ 1 ]  \quad \langle 1256789 \rangle [ z_3, 1 ] \quad \langle 245679 \rangle [ z_1, 1 ] \nonumber \\ & \langle 13678\hspace{0.8mm}10 \rangle [ 1 ] \quad \langle 13458\hspace{0.8mm}10 \rangle [ 1 ] \quad \langle 134579 \rangle [ 1 ]  \quad \langle 12679\hspace{0.8mm}10 \rangle [ 1 ]  \quad \langle 125679 \rangle [ 1 ]  \nonumber \\ & \langle 123679 \rangle [ 1 ] \quad \langle 15679 \rangle [ 1 ] \quad \langle 15678 \rangle [ 1 ] \quad \langle 13679 \rangle [ 1 ] \quad \langle 1347\hspace{0.8mm}10 \rangle [ 1 ] \nonumber \\
& \langle 2679 \rangle [ 1 ] \quad \langle 167\hspace{0.8mm}10 \rangle [ 1 ] \,,
\end{align}
where the values inside the square brackets represent the irreducible
numerators for the given topology. The numerators are expressed in the
Baikov representation using the variables $z_i$. Here $i$ is an
integer corresponding to the $i$th element of the list
\lsm{Propagators}. The possible values of $i$ are determined by the
uncut propagators, for example for the $\langle
1236789\hspace{0.8mm}10 \rangle$ subdiagram the numerators can be chosen as:
\begin{equation}
\begin{alignedat}{3}
&z_{4}=(l_3)^2\,, \quad &&z_{5}=(l_3+k_1+k_2)^2 \,, \quad &&z_{11}=(l_1+k_4)^2 \,,  \\
&z_{12}=(l_2+k_1)^2 \,, \quad && z_{13}=(l_3+k_1)^2 \,, \quad &&z_{14}=(l_3+k_4)^2 \,, \\
&z_{15}=(l_1+l_2)^2 \,.
\label{ispc}
\end{alignedat}
\end{equation}
Other than the $5$ initial ISPs the two uncut denominators $\{z_4,z_5\}$ can appear as numerators.\\
The reduction is very efficient. This is evidenced in fig.~\ref{ex_rd} which displays results for a variety of diagrams at various loop orders and configurations of internal and external masses\footnote{Part of these results were already known, and our results agree with the literature, see for example refs.~\citep{Meyer:2016slj,Bonciani:2016qxi,DiVita:2014pza}.}. Here, $\blacktriangle$, $\blacksquare$ and \tikz\draw[fill=black] (0,0) circle (.75ex); represent different masses.

\subsection{IBP identities evaluated on their maximal cut}

\Azurite\ can also obtain IBP identities on maximal cuts, both analytically or numerically, using the function \lsm{IntegralRed}.
For example, for the triple-box diagram (Azurite sample code
\ref{tb_def}), the analytic IBP identities on the maximal cut
$D_1=D_2=\ldots =D_{10}=0$ can be obtained.
\begin{MathPlain}[float,floatplacement=H,caption=Sample code for \lsm{IntegralRed}]
IntegralRed[{1, 2, 3, 4, 5, 6, 7, 8, 9, 10}];
\end{MathPlain}
The output is a list whose first element contains the master
integrals and the second one contains the IBP
identities evaluated on the cut, in the form of replacement rules.

Using a similar notation as in eq.~\eqref{mi3b} we can represent the master integrals for this cut as $\{ \text{I}[z_{11}] , \text{I}[z_{13}] , \text{I}[1]\}$. Here the prescription $\text{I}[\text{N}]$ indicates, using again the Baikov representation,
\begin{equation}
\text{I}[\text{N}]= \int \prod_{i=11}^{15} \d z_{i} ~ F^{\frac{d-7}{2}} \text{N}.
\end{equation}
The values of $i$ run over the ISPs of this diagram, which can be read off from
eq.~\eqref{ispc}. For instance, the reduction of $\text{I}[z_{11}^2]$ is then written as
\begin{align}
\text{I}[z_{11}^2]=
\frac{1}{2 \left(-3 + d \right) ^2}& \left(-2 \left( \left(8 - 6 d +
                                     d^2 \right) s - \left(-3 + d
                                     \right) t \right)
                                     \text{I}[z_{11}] \right.
                                     \nonumber \\ &\left. + 2 \left(-4
                                                    + d \right)^2 t ~
                                                    \text{I}[z_{13}]
                                                    + \left(-4 + d
                                                    \right) \left(-2 +
                                                    d \right) s t
                                                    ~\text{I}[1]
                                                    \right) +\ldots.
\end{align}
where $\ldots$ denotes integrals with fewer-than-ten propagators. It
takes about $2.4$ seconds to reduce all numerators up to rank $4$ to the
master integrals, and about $18.0$ seconds to reduce all numerators up to rank $6$ to the
master integrals, on the maximal cut, with the same computer mentioned
in the previous subsection.

\section{Summary and Outlook}
In this paper, we have introduced our new algorithm for finding
bases of loop integrals and its implementation in the package \Azurite. It
constructs the needed integration-by-parts identities on a specific
set of (algorithmically determined) cuts, and constructs identities
where integrals with higher-power propagators are absent by solving
syzygy equations. By making use of further simplifications, involving
adaptive parametrizations of the involved diagrams, using graph theory
tools to find discrete symmetries, finite-field computations and parallel
computations, the
package finds master integrals for two- and three-loop diagrams very
efficiently. Therefore we expect that \Azurite\ will be a very useful
tool for studies of multi-loop scattering amplitudes, for example in
IBP reductions and differential equations. This package can also be used to find the
IBP relations evaluated on their maximal cuts analytically.

There are several directions for developing new versions of
\Azurite. One direction is to write a new syzygy generating code,
based on new developments in computational algebraic geometry such as
Faug\`ere's F5 algorithm \cite{Faugere:2002:NEA:780506.780516}. The goal
is to get the code to produce  a simpler form of syzygy generators,
which would allow speeding up the
search for master integrals.  It will also be very helpful to fully
incorporate the tangent Lie algebra/variety duality \cite{Hauser1993}
for deriving syzygies.
Furthermore, we are working on a public package to produce complete
IBP reductions efficiently, based on the present algorithm to find a
basis of integrals, and on the construction of IBP reductions on cuts
via syzygy computations \cite{Ita:2015tya,Larsen:2015ped}.

\section*{Acknowledgements}
We thank S. Badger, Z. Bern, J. Bosma, L. Dixon, C. Duhr,
H. Frellesvig, J. Henn, H. Ita, H. Johansson, D. Kosower, A. von Manteuffel,
F. Moriello, E. Panzer, C. Papadopoulos, R. Schabinger and M. Zeng for
useful discussions. Especially, we thank S. Badger and H. Ita for
testing our package and for careful reading of our manuscripts during
the draft stage. The research leading to these results has received
funding from the European Union Seventh Framework Programme
(FP7/2007-2013) under grant agreement no. 627521, and Swiss National
Science Foundation (Ambizione grant PZ00P2 161341). The work of AG is
supported by the Knut and Alice Wallenberg Foundation under grant
\#2015-0083. The work of AG and YZ is also partially supported by the
Swiss National Science Foundation through the NCCR SwissMap. The work of KJL is supported by ERC-2014-CoG, Grant number 648630 IQFT.

\appendix

\section{Usage of Azurite}\label{usage_azu}
\subsection{Installation}
To install \Azurite, it is necessary to install the computer algebra systems \mm\ (10.0.0 or more
recent versions) and \Singular\ \cite{DGPS} first. \Singular\ can be downloaded from \url{http: //www.singular.uni-kl.de}.
\Azurite\ version $a$.$b$.$c$ can be downloaded from,
\begin{quotation}
  \noindent \url{https://bitbucket.org/yzhphy/azurite/raw/master/release/Azurite_a.b.c.tar.gz}
\end{quotation}
Here $a$, $b$ and $c$
must be replaced by the corresponding version numbers, for example,
\begin{quotation}
\noindent\url{https://bitbucket.org/yzhphy/azurite/raw/master/release/Azurite_1.1.0.tar.gz}.
\end{quotation}

After extracting the tar file \verb&Azurite_a.b.c.tar.gz&, there will be a
directory \verb&Azurite_a.b.c& which consists of the sub-directories
\verb&code&, \verb&examples& and \verb&manual&. The main package file
\verb&Azurite.wl& is located in \verb&code&. \verb&examples& contains
examples while \verb&manual& contains a manual of \Azurite\ in \mm\
notebook format.

A directory for temporary files must be created by the user.

\subsection{Commands and Options}
\subsubsection{Path setup}
The paths of \Azurite\ for temporary files and \Singular\ binary file
are set up as follows, in \mm\ code. For example,
\begin{MathPlain}
<< "/MyPathforAzurite/Azurite/code/Azurite.wl";
TemporaryDirectory = "/MyPathforTemporaryFiles/";
SingularDirectory = "/Applications/Singular.app/Contents/bin/";
\end{MathPlain}
Here \lsm{TemporaryDirectory} denotes the directory of temporary files, while the variable \lsm{SingularDirectory} denotes the directory of the \Singular\ binary file which depends on the operating system.

\subsubsection{Kinematics and loop structure information}
The loop structure and kinematics information should be added after the
path set-up section. The names of loop momenta and external momenta
are declared in \lsm{LoopMomenta} and \lsm{ExternalMomenta}, respectively. Inverse
propagators, kinematics and numerical values of external invariants are listed in
\lsm{Propagators}, \lsm{Kinematics} and \lsm{Numerics}
respectively. The command \lsm{Preparation[]} finds the Baikov
representation.

For example, the input for the pentagon-box in eq. \eqref{pentabox_invProp}
is,
\begin{MathPlain}
LoopMomenta = {l1, l2};
ExternalMomenta = {k1, k2, k3, k4};
Propagators = {l1^2, (l1 - k1)^2, (l1 - k1 - k2)^2, (l1 - k1 - k2 - k3)^2, (l2 + k1 + k2 + k3)^2, (l2 + k1 + k2 + k3 + k4)^2,  l2^2, (l1  + l2)^2, (l1 + k4)^2, (l2 + k1)^2, (l2 + k2)^2};
Kinematics = {k1^2 -> 0, k2^2 -> 0, k3^2 -> 0, k4^2 -> 0,  k1 k2 -> s12/2, k1 k3 -> s13/2, k1 k4 -> s14/2, k2 k3 -> s23/2, k2 k4 -> s24/2, k3 k4 -> (-s12 - s13 - s14 - s23 - s24)/2};
Numerics = {s12 -> 1, s13 -> 7, s14 -> 5, s23 -> 17, s24 -> 23};
Preparation[]
\end{MathPlain}
We have the following requirements,
\begin{itemize}
\item Only linearly independent external momenta can appear in
  \lsm{ExternalMomenta} and \lsm{Kinematics}. For example, we do not
  have $k_5$ in the input.
\item The numerical input \lsm{Numerics} is necessary and the numerical values for
  kinematic variables should be
  generic. For example, the following input should be avoided, as the external mass $m_1$ is set to a non-generic value,
\begin{MathPlain}
Kinematics={k1^2 -> m1^2, k2^2 -> 0, k4^2 -> 0, k1 k2 -> 1/2 (-m1^2 + s), k2 k4 -> 1/2 (m1^2 - s - t), k1 k4 -> 1/2 (-m1^2 + t)};
Numerics={s -> 1, t -> 3, m1-> 0};
\end{MathPlain}
From the first line, \Azurite\ will take $k_1$ to be massive for
deriving the Baikov representation. However, \lsm{m1->0} in second line
may make the obtained Baikov representation singular. The correct input
for a massless $k_1$ is,
\begin{MathPlain}
Kinematics={k1^2 -> m1^2, k2^2 -> 0, k4^2 -> 0, k1 k2 -> 1/2 (-m1^2 + s), k2 k4 -> 1/2 (m1^2 - s - t), k1 k4 -> 1/2 (-m1^2 + t)}/.m1->0;
Numerics={s -> 1, t -> 3};
\end{MathPlain}
\item Irreducible scalar products should be added to
  \lsm{Propagators}. The goal is to ensure that the elements in \lsm{Propagators}
  independently span the space of scalar products formed out of $l_i \cdot l_j$
  and $l_i \cdot k_j$. (The $l_i$ are the loop momenta while the $k_j$ are the
  independent external momenta.)

\end{itemize}

\subsubsection{Associated graphs and their discrete symmetries}
After the preparation, \Azurite\ can find graphs and symmetries via the
graph functions of \mm.  For example, with the inverse propagators
given in eq. \eqref{pentabox_invProp},
the graphs of $\la 12345678 \ra$ and $\la 145678 \ra$ are obtained by calling,
\begin{MathPlain}
FeynmanGraph[{1, 2, 3, 4, 5, 6, 7, 8}]
FeynmanGraph[{1, 4, 5, 6, 7, 8}]
\end{MathPlain}
\lsm{FeynmanGraph} has several options, including,
\begin{enumerate}
\item \lsm{DiagramExtendedOutput} with the
default value \lsm{False}. If its
value is \lsm{True}, then propagator labels will appear on the
corresponding internal lines.\footnote{Versions 10 and 11 of \mm\, have
  a problem in labelling  edges of multi graphs: the multiple edges cannot be
  distinctly labelled. We expect that this issue will be solved in
  future versions of \mm.}
\item \lsm{FetchCachedGraphInfo} with the default value \lsm{True}.
To speed up the drawing of graphs, it is advantageous to first store the input
diagram in RAM. This is achieved by calling \lsm{DiagramCache}, for
example,
\begin{MathPlain}
  DiagramCache[{1, 2, 3, 4, 5, 6, 7, 8}];
\end{MathPlain}
Then, provided \lsm{FetchCachedGraphInfo} has the value \lsm{True},
for any subdiagram of $\la 1 2 3 4 5 6 7 8\ra$,
\Azurite\ will simply pinch propagators to obtain the graph, without
running the backtracking algorithm again.
\end{enumerate}

\Azurite\ also finds the discrete symmetries of a given
graph. \lsm{PropagatorSymmetry[index]}
provides the permutation symmetry of propagators.
For example, with the inverse propagators given in eq.
\eqref{pentabox_invProp}, the (physical) symmetry group of diagram $\la 145678
\ra$ is given by \lsm{PropagatorSymmetry[\{1,4,5,6,7,8\}]}. The output is,
\begin{MathPlain}
  {{z[1] -> z[1], z[4] -> z[4], z[5] -> z[5], z[6] -> z[6], z[7] -> z[7], z[8] -> z[8]}, {z[1] -> z[4], z[4] -> z[1], z[5] -> z[7], z[6] -> z[6], z[7] -> z[5], z[8] -> z[8]}}
\end{MathPlain}
where \lsm{z[i]} denotes the Baikov variable $z_i$, namely, the
$i$th propagator.

On the other hand, the action of the symmetries on the momenta can be obtained by \Azurite's \lsm{MomentaSymmetry[index]}. For example, the output of \lsm{MomentaSymmetry[\{1,4,5,6,7,8\}]} reads,
\begin{MathPlain}
{{k4 -> k4, -k1 - k2 - k3 - k4 -> -k1 - k2 - k3 - k4, l1 -> l1, l2 -> l2}, {k4 -> -k1 - k2 - k3 - k4, -k1 - k2 - k3 - k4 -> k4, l1 -> k1 + k2 + k3 - l1, l2 -> -k1 - k2 - k3 - l2}}
\end{MathPlain}
The non-trivial element is  the affine transformation given in eq.
\eqref{145678_Affine}. (Note that \lsm{-k1 - k2 - k3 - k4} means
\lsm{k5}, since the latter  is a linearly dependent momentum.)

\subsubsection{Master integrals}\label{appendix_mi}
\lsm{DiagramAnalysis[index]} provides the list of basis integrals for a given
diagram, without considering its subdiagrams. It is useful for studying
an individual diagram in detail. For example, with the inverse propagators given in eq.
\eqref{pentabox_invProp},
\lsm{DiagramAnalysis[\{1,2,3,4,5,6,7,8\}]} finds the master integrals
of $\la 12345678\ra$. The output is,
\begin{MathPlain}
  {z[10], z[9], 1}
\end{MathPlain}
which means that there are three master integrals supported on the maximal
cut $z_1 = z_2 = ... = z_8 = 0$. They are integrals with numerators $z_{10}=(l_2+k_1)^2$,
$z_9=(l_1+k_4)^2$ and $1$. Similarly,
\lsm{DiagramAnalysis[\{1,4,5,6,7,8\}]} gives \lsm{\{\}}, which means
that this diagram has no master integrals which are supported on the maximal cut $z_1 = z_4 = z_5 = z_6 = z_7 = z_8 = 0$. \lsm{DiagramAnalysis} has the following options,
\begin{itemize}
\item \lsm{NumericMode} with the default value \lsm{True}. This
  determines if the computation is carried out numerically.
\item \lsm{Characteristic} with the default value $0$. This is the
  characteristic of the number field, which can be chosen as either a
  prime number
  $p$, or $0$. In the former case, the finite field $\mathbb Z/p\mathbb
  Z$ is used, while in the latter case the field of rational numbers $\mathbb Q$ is
  used.
\item \lsm{NumericD} with the default value \lsm{Null}. When this
  value is a number, then the spacetime dimension will be set to this numerical value. Note that only rational non-integer values can be
  used.
\item \lsm{WorkingPower} with the default value $4$. This is the
  degree limit for the numerators appearing in the independent IBP
  identities, after Gaussian elimination. In
  general, to get the integral basis, we do not need to reduce all
  renormalizable terms by IBP identities.
\item \lsm{HighestPower} with the default value $4$. Occasionally, to get
  all IBP identities up to the degree specified by \lsm{WorkingPower}, we need
  IBP identities with the degrees higher than \lsm{WorkingPower}. Otherwise, the
  output basis may be redundant and contain integrals with the degree
  exactly the same as
  \lsm{WorkingPower}. \lsm{HighestPower} sets the limit for IBP identities in the
  intermediate steps. \lsm{HighestPower} should be greater than or
  than \lsm{WorkingPower}.
\item \lsm{Symmetry} with the default value \lsm{True}. It determines
  if symmetries are used for the integral reduction.
\item
\lsm{WatchingMode} with the default value \lsm{False}. If it is set to
be \lsm{True},
then the intermediate steps of the computations are printed.
\end{itemize}

\lsm{DiagramAnalysis} applies adaptive parametrizations.
Hence if several external lines attach to one vertex, or if the
diagram is factorizable, it will automatically determine a new list of
ISPs. If the adaptive parametrization is used, the output may contain expressions \lsm{mp2[...]}
which denotes Minkowski scalar products $(...)^2$. Scalar products can be expressed as a function of the original
propagators, via \lsm{SPExpand[]}. For example,
\begin{MathPlain}
SPExpand[mp2[-k1 - k2 - k3 - k4 + l1]]
\end{MathPlain}
gives the output,
\begin{MathPlain}
-s12 - s13 - s23 + z[1] + z[4] - 2 z[9]
\end{MathPlain}
which means $(l_1+k_5)^2=-s_{12}-s_{13}-s_{23}+D_1+D_4-2D_9$.

Note that \lsm{DiagramAnalysis} considers a diagram individually,
and symmetries between different diagrams are ignored. For example,
for the inverse propagators given in eq. \eqref{pentabox_invProp}, \lsm{DiagramAnalysis} determines both
$\la 158\ra[1]$ and $\la 478\ra[1]$ as master integrals. However, they
are equal by a discrete symmetry.

On the other hand, \lsm{FindAllMIs[index]} finds all master integrals
within a diagram and all of its subdiagrams. It first finds the
symmetries between different diagrams, and then determines the
candidate diagrams for the search of master integrals. For example, all
master integrals, including subdiagrams, for the inverse propagators
given in eq. \eqref{pentabox_invProp},
can be found by calling \lsm{FindAllMIs[\{1,2,3,4,5,6,7,8\}]}. During the computation of \lsm{FindAllMIs}, the obtained master
integrals are printed in the following format,
\begin{MathPlain}
{1,2,3,4,5,6,7,8}      {z[10],z[9],1}
{2,3,4,5,6,7,8}      {z[9],z[1],1}
...
\end{MathPlain}
For each line the first entry is the list of propagators of a diagram, while the second entry
is the list of numerators for master integrals of this topology. When the computation
has finished, the total time used is also displayed. The output of
\lsm{FindAllMIs} is a list which consists of items whose first element is the
diagram index, and the second item is the list of numerators of master
integrals.

After calling \lsm{FindAllMIs}, the associated diagrams of master integrals can be
obtained and displayed by calling,
\begin{MathPlain}
MIList = FindAllMIs[{1, 2, 3, 4, 5, 6, 7, 8}];
FeynmanGraph[#[[1]],DiagramExtendedOutput -> True] & /@ MIList
\end{MathPlain}

Most options of \lsm{FindAllMIs} are the same as those of
\lsm{DiagramAnalysis}. However to speed up the search process, some
default values are different:
\begin{MathPlain}
{NumericMode->True,Characteristic->9001,NumericD->1138/17,HighestPower->4,WorkingPower->3,
WatchingMode->False,Symmetry->True, GlobalSymmetry->True, ParallelMode->True}
\end{MathPlain}
Here \lsm{GlobalSymmetry} is a special option which determines whether the
symmetries between different diagrams are used. \lsm{ParallelMode} is
the option which indicates if the parallel computation is used. If
its value is \lsm{True}, then the sub-diagrams are assigned to several
processors, and the integral basis searching can be significantly sped
up.

\subsubsection{Analytic IBP identities evaluated on their maximal cut}
Analytic (or numerical) IBP identities evaluated on their maximal cut
can be obtained by the
\lsm{IntegralRed} command. For example, for the triple-box diagram
(shown in Azurite code sample \ref{tb_def}), the IBP identities on the maximal ($10$-propagator) cut can obtained via,
\begin{MathPlain}
{MIs,IBP}=IntegralRed[{1, 2, 3, 4, 5, 6, 7, 8, 9, 10}];
\end{MathPlain}
where the variable \lsm{MIs} contains master integrals and \lsm{IBP} contains
IBP identities evaluated on their maximal cut, in the form of replacement rules. The reduction of a specific integral
can now be obtained,
\begin{MathPlain}
Int[1, 1, 1, 1, 1, 1, 1, 1, 1, 1, -3, 0, 0, 0, 0] /. IBP
\end{MathPlain}
where $\mathrm{Int}[a_1 , \ldots, a_{k} ,\ldots a_{n_\text{SP}}]$ denotes the integral with the integrand,
\begin{equation}
\label{integral_index}
\frac{D_{k+1}^{-a_{k+1}} \cdots D_{n_\text{SP}}^{-a_{n_\text{SP}}}}{D_1^{a_1} \cdots D_k^{a_k}}.
\end{equation}
where $D_{k+1},\ldots D_{n_\text{SP}}$ are irreducible scalar products.

The options for \lsm{IntegralRed} are similar to those of
\lsm{DiagramAnalysis},
except that the default values are for analytic computation:
\begin{MathPlain}
{NumericMode->False,Characteristic->0,NumericD->Null,HighestPower->4,WorkingPower->4, WatchingMode->False,Symmetry->True}
\end{MathPlain}
Note that \lsm{IntegralRed} does not use adaptive
parametrization. Hence for \lsm{IntegralRed}, the kinematic input must correspond
to a non-factorizable diagram whose external lines attach to distinct
vertices.

\section{Integrals with squared propagators in Baikov representation on maximal cuts}
In this paper, we mainly discuss integrals without squared propagators. However, integrals with squared propagators do appear in various contexts, for example, importantly, in the context of differential equations \cite{Kotikov:1990kg,Kotikov:1991pm,Bern:1993kr,Remiddi:1997ny,Gehrmann:1999as,Ablinger:2015tua,Henn:2013pwa}. To explain the relation between integrals with and without squared propagators, in this section we derive the form of squared-propagator integrals in their Baikov representation on maximal cuts.

Recall that in eq.~\eqref{Baikov_mc}, the maximal-cut form of an integral without squared propagators is obtained by simply setting $z_1$, ..., $z_m$ to zero in the integrand. If a propagator $z_j$ is squared ($1\leq j\leq m$), then a residue computation is necessary to obtain the Baikov representation on the maximal cut. In the notation of eq.~\eqref{Baikov_mc}, the integral with the integrand $N/(D_1\cdots D_j^2\cdots D_m)$, evaluated on the maximal cut is proportional to,
\begin{align}
  &\hspace{2mm}
 \int \d z_{m+1} \cdots \d z_{n_\text{SP}'} \oint_{\mathcal C_j} \d z_j \frac{1}{z_j^2} F(0,\ldots,z_j \ldots ,0,z_{m+1},\ldots , z_{n_\text{SP}'})^{\frac{D-h}{2}}\nn\\
  &\hspace{20mm} \times N(0, \ldots,z_j,\ldots,0,z_{m+1},\ldots , z_{n_\text{SP}'}) \,,
\label{Baikov_sq_mc}
\end{align}
in Baikov representation. Here $\mathcal C_j$ is a small contour around the point $z_j=0$. Furthermore, assume that the integrand has been reduced so that the numerator $N$ is independent of $z_j$. After evaluating this residue, the maximal-cut form reads,
\begin{gather}
   \int \d z_{m+1} \cdots \d z_{n_\text{SP}'} F(0,\ldots,0,z_{m+1},\ldots , z_{n_\text{SP}'})^{\frac{D-h-2}{2}}\nn\\
  \times \frac{D-h}{2} \frac{\partial F}{\partial z_j}(0,\ldots,0,z_{m+1},\ldots , z_{n_\text{SP}'}) N(0, \ldots,0,z_{m+1},\ldots , z_{n_\text{SP}'}) \,.
\end{gather}
Therefore a squared-propagator integral evaluated on its maximal cut, in its Baikov representation, is equivalent to a $(D-2)$-dimensional integral without squared propagators. So by dimension-shift identities and IBP identities, a $D$-dimensional integral with squared propagators equals a linear combination of $D$-dimensional integrals without squared propagators.

As an example, consider the two-loop four-point massless double-box diagram with inverse propagators:
\begin{equation}
\begin{alignedat}{4}
  & D_1 = l_1^2\,,              \quad  && D_2 = (l_1 - k_1)^2\,,  \quad  && D_3 = (l_1 - K_{12})^2\,,  \quad  && D_4 = (l_1 + K_{12})^2 \,, \\
  & D_5 = (l_2 - k_4)^2\,,  \quad  && D_6 = l_2 ^2\,,   \quad  && D_7 = (l_1+l_2)^2\,,
\label{pentabox_invProp}
\end{alignedat}
\end{equation}
where $k_1^2=k_2^2=k_3^2=k_4^2=0$, $(k_1 +k_2)^2=s$ and $(k_1 +k_4)^2=t$. As in eq.~\eqref{integral_index}, we define
\begin{equation}
  \label{eq:12}
  I[m_1,\ldots m_9;D]=\int\frac{\d^D l_1}{i \pi^{D/2}}\frac{\d^D l_2}{i \pi^{D/2}}\frac{(l_1+k_4)^{-m_8} (l_2+k_1)^{-m_9}}{D_1^{m_1} \cdots D_7^{m_7}}\,.
\end{equation}
Define Baikov variables as, $z_i\equiv D_i$, $i=1,\ldots, 7$, $z_8\equiv(l_1+k_4)^2$ and $z_9\equiv(l_2+k_1)^2$. The Baikov representation is,
\begin{equation}
  \label{eq:13}
   I[m_1,\ldots m_9;D]=C(D) \int \prod_{i=1}^9 \d z_i F(z)^{\frac{D-6}{2}} \frac{z_8^{-m_8}z_9^{-m_9}}{z_1^{m_1}\ldots z_7^{m_7}}\,,
\end{equation}
where $C(D)$ is a dimension-dependent prefactor. Now consider the maximal cut $z_1=\ldots =z_7=0$. For example,
\begin{align}
  I[1,1,1,1,1,1,2,0,0;D]=C(D) \int \d z_8 \d z_9 \frac{D-6}{2} \bigg(\frac{\partial F}{\partial z_7} F(z) ^{\frac{D-8}{2}} \bigg)\bigg|_{z_1=\ldots =z_7=0}.
\end{align}
Hence on the maximal cut, $I[1,1,1,1,1,1,2,0,0;D]$ equals a $(D-2)$-dimensional integral with the numerator $\partial F/\partial z_7$, but without squared propagators. Using $(D-2)$-dimensional IBP identities, we obtain,
\begin{align}
  I[1,1,1,1,1,1,2,0,0;D]=\frac{C(D)}{C(D-2)} ( a_1 B_1[D-2]+a_2 B_2[D-2])\,,
\label{sq}
\end{align}
where
\begin{align}
  B_1[D]&\equiv I[1,1,1,1,1,1,1,0,0;D-2]\,,\nn\\
B_2[D]&\equiv I[1,1,1,1,1,1,1,-1,0;D-2]\,,
\end{align}
are the two master integrals of the double-box topology. The coefficients are $a_1=(D-6)s^2/(16(s+t))$ and $a_2=-(D-6)s(3s+2t)/(16t(s+t))$. Here $\ldots$ denotes integrals with fewer-than-seven propagators.

On the other hand,
\begin{align}
  I[m_1,\ldots m_9;D]=C(D) \int \prod_{i=1}^9 \d z_i F(z)^{\frac{D-8}{2}} \frac{z_8^{-m_8}z_9^{-m_9} F(z)}{z_1^{m_1}\ldots z_7^{m_7}}\,.
\end{align}
which implies dimension-shift identities. Again using $(D-2)$-dimensional IBP identities,
\begin{align}
   B_i[D]=\frac{C(D)}{C(D-2)} \big( T_{1i} B_1[D-2]+T_{2i} B_2[D-2]) + \ldots\,,
\label{dim_shift}
\end{align}
where $i=1,2$. Now compare eqs.~\eqref{sq} and \eqref{dim_shift}, and define
\begin{equation}
  \label{eq:15}
 \left(
  \begin{array}{c}
    c_1\\
    c_2
  \end{array}
\right)\equiv
  \left(
  \begin{array}{cc}
    T_{11}&T_{12}\\
    T_{21}&T_{22}
  \end{array}
\right)^{-1}
  \left(
  \begin{array}{c}
    a_1\\
    a_2
  \end{array}
\right)\,.
\end{equation}
Then, on the maximal cut, the squared-propagator integral is related to integrals without squared propagators as $I[1,1,1,1,1,1,2,0,0;D]=c_1 B_1[D]+c_2 B_2[D]+\ldots$. Here,
\begin{equation}
  \label{eq:11}
  c_1=-\frac{(D-5)(3D-14)}{(D-6)t}\,,\quad c_2=-\frac{2(D-5)(D-4)}{(D-6)st}\,.
\end{equation}
Note that the explicit form of $C(D)$ is not needed for deriving these coefficients.

This representation of integrals with squared propagators on maximal cuts clearly generalizes to integrals with squared propagators on non-maximal cuts.





\bibliographystyle{elsarticle-num}
\bibliography{azurite}

\begin{thebibliography}{10}
\expandafter\ifx\csname url\endcsname\relax
  \def\url#1{\texttt{#1}}\fi
\expandafter\ifx\csname urlprefix\endcsname\relax\def\urlprefix{URL }\fi
\expandafter\ifx\csname href\endcsname\relax
  \def\href#1#2{#2} \def\path#1{#1}\fi

\bibitem{Tkachov:1981wb}
F.~Tkachov, {A Theorem on Analytical Calculability of Four Loop Renormalization
  Group Functions}, Phys.Lett. B100 (1981) 65--68.
\newblock \href {http://dx.doi.org/10.1016/0370-2693(81)90288-4}
  {\path{doi:10.1016/0370-2693(81)90288-4}}.

\bibitem{Chetyrkin:1981qh}
K.~Chetyrkin, F.~Tkachov, {Integration by Parts: The Algorithm to Calculate
  beta Functions in 4 Loops}, Nucl.Phys. B192 (1981) 159--204.
\newblock \href {http://dx.doi.org/10.1016/0550-3213(81)90199-1}
  {\path{doi:10.1016/0550-3213(81)90199-1}}.

\bibitem{Smirnov:2010hn}
A.~V. Smirnov, A.~V. Petukhov, {The Number of Master Integrals is Finite},
  Lett. Math. Phys. 97 (2011) 37--44.
\newblock \href {http://arxiv.org/abs/1004.4199} {\path{arXiv:1004.4199}},
  \href {http://dx.doi.org/10.1007/s11005-010-0450-0}
  {\path{doi:10.1007/s11005-010-0450-0}}.

\bibitem{Laporta:2001dd}
S.~Laporta, {High precision calculation of multiloop Feynman integrals by
  difference equations}, Int.J.Mod.Phys. A15 (2000) 5087--5159.
\newblock \href {http://arxiv.org/abs/hep-ph/0102033}
  {\path{arXiv:hep-ph/0102033}}, \href
  {http://dx.doi.org/10.1016/S0217-751X(00)00215-7}
  {\path{doi:10.1016/S0217-751X(00)00215-7}}.

\bibitem{Laporta:2000dc}
S.~Laporta, {Calculation of master integrals by difference equations}, Phys.
  Lett. B504 (2001) 188--194.
\newblock \href {http://arxiv.org/abs/hep-ph/0102032}
  {\path{arXiv:hep-ph/0102032}}, \href
  {http://dx.doi.org/10.1016/S0370-2693(01)00256-8}
  {\path{doi:10.1016/S0370-2693(01)00256-8}}.

\bibitem{Anastasiou:2004vj}
C.~Anastasiou, A.~Lazopoulos, {Automatic integral reduction for higher order
  perturbative calculations}, JHEP 0407 (2004) 046.
\newblock \href {http://arxiv.org/abs/hep-ph/0404258}
  {\path{arXiv:hep-ph/0404258}}, \href
  {http://dx.doi.org/10.1088/1126-6708/2004/07/046}
  {\path{doi:10.1088/1126-6708/2004/07/046}}.

\bibitem{Smirnov:2008iw}
A.~Smirnov, {Algorithm FIRE -- Feynman Integral REduction}, JHEP 0810 (2008)
  107.
\newblock \href {http://arxiv.org/abs/0807.3243} {\path{arXiv:0807.3243}},
  \href {http://dx.doi.org/10.1088/1126-6708/2008/10/107}
  {\path{doi:10.1088/1126-6708/2008/10/107}}.

\bibitem{Smirnov:2014hma}
A.~V. Smirnov, {FIRE5: a C++ implementation of Feynman Integral REduction},
  Comput. Phys. Commun. 189 (2015) 182--191.
\newblock \href {http://arxiv.org/abs/1408.2372} {\path{arXiv:1408.2372}},
  \href {http://dx.doi.org/10.1016/j.cpc.2014.11.024}
  {\path{doi:10.1016/j.cpc.2014.11.024}}.

\bibitem{Studerus:2009ye}
C.~Studerus, {Reduze-Feynman Integral Reduction in C++}, Comput.Phys.Commun.
  181 (2010) 1293--1300.
\newblock \href {http://arxiv.org/abs/0912.2546} {\path{arXiv:0912.2546}},
  \href {http://dx.doi.org/10.1016/j.cpc.2010.03.012}
  {\path{doi:10.1016/j.cpc.2010.03.012}}.

\bibitem{vonManteuffel:2012yz}
A.~von Manteuffel, C.~Studerus, {Reduze 2 - Distributed Feynman Integral
  Reduction}\href {http://arxiv.org/abs/1201.4330} {\path{arXiv:1201.4330}}.

\bibitem{Lee:2012cn}
R.~N. Lee, {Presenting LiteRed: a tool for the Loop InTEgrals REDuction}\href
  {http://arxiv.org/abs/1212.2685} {\path{arXiv:1212.2685}}.

\bibitem{vonManteuffel:2014ixa}
A.~von Manteuffel, R.~M. Schabinger, {A novel approach to integration by parts
  reduction}, Phys. Lett. B744 (2015) 101--104.
\newblock \href {http://arxiv.org/abs/1406.4513} {\path{arXiv:1406.4513}},
  \href {http://dx.doi.org/10.1016/j.physletb.2015.03.029}
  {\path{doi:10.1016/j.physletb.2015.03.029}}.

\bibitem{vonManteuffel:2015gxa}
A.~von Manteuffel, E.~Panzer, R.~M. Schabinger, {On the Computation of Form
  Factors in Massless QCD with Finite Master Integrals}, Phys. Rev. D93~(12)
  (2016) 125014.
\newblock \href {http://arxiv.org/abs/1510.06758} {\path{arXiv:1510.06758}},
  \href {http://dx.doi.org/10.1103/PhysRevD.93.125014}
  {\path{doi:10.1103/PhysRevD.93.125014}}.

\bibitem{vonManteuffel:2016xki}
A.~von Manteuffel, R.~M. Schabinger, {Quark and gluon form factors to four loop
  order in QCD: the $N_f^3$ contributions}\href
  {http://arxiv.org/abs/1611.00795} {\path{arXiv:1611.00795}}.

\bibitem{Peraro:2016wsq}
T.~Peraro, {Scattering amplitudes over finite fields and multivariate
  functional reconstruction}, JHEP 12 (2016) 030.
\newblock \href {http://arxiv.org/abs/1608.01902} {\path{arXiv:1608.01902}},
  \href {http://dx.doi.org/10.1007/JHEP12(2016)030}
  {\path{doi:10.1007/JHEP12(2016)030}}.

\bibitem{Gluza:2010ws}
J.~Gluza, K.~Kajda, D.~A. Kosower, {Towards a Basis for Planar Two-Loop
  Integrals}, Phys.Rev. D83 (2011) 045012.
\newblock \href {http://arxiv.org/abs/1009.0472} {\path{arXiv:1009.0472}},
  \href {http://dx.doi.org/10.1103/PhysRevD.83.045012}
  {\path{doi:10.1103/PhysRevD.83.045012}}.

\bibitem{Schabinger:2011dz}
R.~M. Schabinger, {A New Algorithm For The Generation Of Unitarity-Compatible
  Integration By Parts Relations}, JHEP 1201 (2012) 077.
\newblock \href {http://arxiv.org/abs/1111.4220} {\path{arXiv:1111.4220}},
  \href {http://dx.doi.org/10.1007/JHEP01(2012)077}
  {\path{doi:10.1007/JHEP01(2012)077}}.

\bibitem{Kotikov:1990kg}
A.~V. Kotikov, {Differential equations method: New technique for massive
  Feynman diagrams calculation}, Phys. Lett. B254 (1991) 158--164.
\newblock \href {http://dx.doi.org/10.1016/0370-2693(91)90413-K}
  {\path{doi:10.1016/0370-2693(91)90413-K}}.

\bibitem{Kotikov:1991pm}
A.~V. Kotikov, {Differential equation method: The Calculation of N point
  Feynman diagrams}, Phys. Lett. B267 (1991) 123--127, [Erratum: Phys.
  Lett.B295,409(1992)].
\newblock \href {http://dx.doi.org/10.1016/0370-2693(91)90536-Y,
  10.1016/0370-2693(92)91582-T} {\path{doi:10.1016/0370-2693(91)90536-Y,
  10.1016/0370-2693(92)91582-T}}.

\bibitem{Bern:1993kr}
Z.~Bern, L.~J. Dixon, D.~A. Kosower, {Dimensionally regulated pentagon
  integrals}, Nucl. Phys. B412 (1994) 751--816.
\newblock \href {http://arxiv.org/abs/hep-ph/9306240}
  {\path{arXiv:hep-ph/9306240}}, \href
  {http://dx.doi.org/10.1016/0550-3213(94)90398-0}
  {\path{doi:10.1016/0550-3213(94)90398-0}}.

\bibitem{Remiddi:1997ny}
E.~Remiddi, {Differential equations for Feynman graph amplitudes}, Nuovo Cim.
  A110 (1997) 1435--1452.
\newblock \href {http://arxiv.org/abs/hep-th/9711188}
  {\path{arXiv:hep-th/9711188}}.

\bibitem{Gehrmann:1999as}
T.~Gehrmann, E.~Remiddi, {Differential equations for two loop four point
  functions}, Nucl.Phys. B580 (2000) 485--518.
\newblock \href {http://arxiv.org/abs/hep-ph/9912329}
  {\path{arXiv:hep-ph/9912329}}, \href
  {http://dx.doi.org/10.1016/S0550-3213(00)00223-6}
  {\path{doi:10.1016/S0550-3213(00)00223-6}}.

\bibitem{Ablinger:2015tua}
J.~Ablinger, A.~Behring, J.~Blümlein, A.~De~Freitas, A.~von Manteuffel,
  C.~Schneider, {Calculating Three Loop Ladder and V-Topologies for Massive
  Operator Matrix Elements by Computer Algebra}, Comput. Phys. Commun. 202
  (2016) 33--112.
\newblock \href {http://arxiv.org/abs/1509.08324} {\path{arXiv:1509.08324}},
  \href {http://dx.doi.org/10.1016/j.cpc.2016.01.002}
  {\path{doi:10.1016/j.cpc.2016.01.002}}.

\bibitem{Henn:2013pwa}
J.~M. Henn, {Multiloop integrals in dimensional regularization made simple},
  Phys. Rev. Lett. 110 (2013) 251601.
\newblock \href {http://arxiv.org/abs/1304.1806} {\path{arXiv:1304.1806}},
  \href {http://dx.doi.org/10.1103/PhysRevLett.110.251601}
  {\path{doi:10.1103/PhysRevLett.110.251601}}.

\bibitem{Lee:2014ioa}
R.~N. Lee, {Reducing differential equations for multiloop master integrals},
  JHEP 04 (2015) 108.
\newblock \href {http://arxiv.org/abs/1411.0911} {\path{arXiv:1411.0911}},
  \href {http://dx.doi.org/10.1007/JHEP04(2015)108}
  {\path{doi:10.1007/JHEP04(2015)108}}.

\bibitem{Meyer:2016slj}
C.~Meyer, {Transforming differential equations of multi-loop Feynman integrals
  into canonical form}\href {http://arxiv.org/abs/1611.01087}
  {\path{arXiv:1611.01087}}.

\bibitem{Remiddi:2016gno}
E.~Remiddi, L.~Tancredi, {Differential equations and dispersion relations for
  Feynman amplitudes. The two-loop massive sunrise and the kite integral},
  Nucl. Phys. B907 (2016) 400--444.
\newblock \href {http://arxiv.org/abs/1602.01481} {\path{arXiv:1602.01481}},
  \href {http://dx.doi.org/10.1016/j.nuclphysb.2016.04.013}
  {\path{doi:10.1016/j.nuclphysb.2016.04.013}}.

\bibitem{Bonciani:2016qxi}
R.~Bonciani, V.~Del~Duca, H.~Frellesvig, J.~M. Henn, F.~Moriello, V.~A.
  Smirnov, {Two-loop planar master integrals for Higgs$\to 3$ partons with full
  heavy-quark mass dependence}\href {http://arxiv.org/abs/1609.06685}
  {\path{arXiv:1609.06685}}.

\bibitem{Primo:2016ebd}
A.~Primo, L.~Tancredi, {On the maximal cut of Feynman integrals and the
  solution of their differential equations}\href
  {http://arxiv.org/abs/1610.08397} {\path{arXiv:1610.08397}}.

\bibitem{Larsen:2015ped}
K.~J. Larsen, Y.~Zhang, {Integration-by-parts reductions from unitarity cuts
  and algebraic geometry}, Phys. Rev. D93~(4) (2016) 041701.
\newblock \href {http://arxiv.org/abs/1511.01071} {\path{arXiv:1511.01071}},
  \href {http://dx.doi.org/10.1103/PhysRevD.93.041701}
  {\path{doi:10.1103/PhysRevD.93.041701}}.

\bibitem{Ita:2015tya}
H.~Ita, {Two-loop Integrand Decomposition into Master Integrals and Surface
  Terms}, Phys. Rev. D94~(11) (2016) 116015.
\newblock \href {http://arxiv.org/abs/1510.05626} {\path{arXiv:1510.05626}},
  \href {http://dx.doi.org/10.1103/PhysRevD.94.116015}
  {\path{doi:10.1103/PhysRevD.94.116015}}.

\bibitem{DGPS}
W.~Decker, G.-M. Greuel, G.~Pfister, H.~Sch\"onemann, {\sc Singular} {4-0-2}
  --- {A} computer algebra system for polynomial computations,
  \url{http://www.singular.uni-kl.de} (2015).

\bibitem{Lee:2013hzt}
R.~N. Lee, A.~A. Pomeransky, {Critical points and number of master integrals},
  JHEP 11 (2013) 165.
\newblock \href {http://arxiv.org/abs/1308.6676} {\path{arXiv:1308.6676}},
  \href {http://dx.doi.org/10.1007/JHEP11(2013)165}
  {\path{doi:10.1007/JHEP11(2013)165}}.

\bibitem{Ossola:2006us}
G.~Ossola, C.~G. Papadopoulos, R.~Pittau, {Reducing full one-loop amplitudes to
  scalar integrals at the integrand level}, Nucl.Phys. B763 (2007) 147--169.
\newblock \href {http://arxiv.org/abs/hep-ph/0609007}
  {\path{arXiv:hep-ph/0609007}}, \href
  {http://dx.doi.org/10.1016/j.nuclphysb.2006.11.012}
  {\path{doi:10.1016/j.nuclphysb.2006.11.012}}.

\bibitem{Ossola:2007ax}
G.~Ossola, C.~G. Papadopoulos, R.~Pittau, {CutTools: A Program implementing the
  OPP reduction method to compute one-loop amplitudes}, JHEP 0803 (2008) 042.
\newblock \href {http://arxiv.org/abs/0711.3596} {\path{arXiv:0711.3596}},
  \href {http://dx.doi.org/10.1088/1126-6708/2008/03/042}
  {\path{doi:10.1088/1126-6708/2008/03/042}}.

\bibitem{Ellis:2011cr}
R.~Ellis, Z.~Kunszt, K.~Melnikov, G.~Zanderighi, {One-loop calculations in
  quantum field theory: from Feynman diagrams to unitarity cuts}\href
  {http://arxiv.org/abs/1105.4319} {\path{arXiv:1105.4319}}.

\bibitem{Ellis:2007br}
R.~Ellis, W.~Giele, Z.~Kunszt, {A Numerical Unitarity Formalism for Evaluating
  One-Loop Amplitudes}, JHEP 0803 (2008) 003.
\newblock \href {http://arxiv.org/abs/0708.2398} {\path{arXiv:0708.2398}},
  \href {http://dx.doi.org/10.1088/1126-6708/2008/03/003}
  {\path{doi:10.1088/1126-6708/2008/03/003}}.

\bibitem{Mastrolia:2011pr}
P.~Mastrolia, G.~Ossola, {On the Integrand-Reduction Method for Two-Loop
  Scattering Amplitudes}, JHEP 1111 (2011) 014.
\newblock \href {http://arxiv.org/abs/1107.6041} {\path{arXiv:1107.6041}},
  \href {http://dx.doi.org/10.1007/JHEP11(2011)014}
  {\path{doi:10.1007/JHEP11(2011)014}}.

\bibitem{Badger:2012dp}
S.~Badger, H.~Frellesvig, Y.~Zhang, {Hepta-Cuts of Two-Loop Scattering
  Amplitudes}, JHEP 1204 (2012) 055.
\newblock \href {http://arxiv.org/abs/1202.2019} {\path{arXiv:1202.2019}},
  \href {http://dx.doi.org/10.1007/JHEP04(2012)055}
  {\path{doi:10.1007/JHEP04(2012)055}}.

\bibitem{Zhang:2012ce}
Y.~Zhang, {Integrand-Level Reduction of Loop Amplitudes by Computational
  Algebraic Geometry Methods}, JHEP 1209 (2012) 042.
\newblock \href {http://arxiv.org/abs/1205.5707} {\path{arXiv:1205.5707}},
  \href {http://dx.doi.org/10.1007/JHEP09(2012)042}
  {\path{doi:10.1007/JHEP09(2012)042}}.

\bibitem{Mastrolia:2012an}
P.~Mastrolia, E.~Mirabella, G.~Ossola, T.~Peraro, {Scattering Amplitudes from
  Multivariate Polynomial Division}, Phys.Lett. B718 (2012) 173--177.
\newblock \href {http://arxiv.org/abs/1205.7087} {\path{arXiv:1205.7087}},
  \href {http://dx.doi.org/10.1016/j.physletb.2012.09.053}
  {\path{doi:10.1016/j.physletb.2012.09.053}}.

\bibitem{Baikov:1996rk}
P.~A. Baikov, {Explicit solutions of the three loop vacuum integral recurrence
  relations}, Phys. Lett. B385 (1996) 404--410.
\newblock \href {http://arxiv.org/abs/hep-ph/9603267}
  {\path{arXiv:hep-ph/9603267}}, \href
  {http://dx.doi.org/10.1016/0370-2693(96)00835-0}
  {\path{doi:10.1016/0370-2693(96)00835-0}}.

\bibitem{vanNeerven:1983vr}
W.~L. van Neerven, J.~A.~M. Vermaseren, {LARGE LOOP INTEGRALS}, Phys. Lett.
  B137 (1984) 241--244.
\newblock \href {http://dx.doi.org/10.1016/0370-2693(84)90237-5}
  {\path{doi:10.1016/0370-2693(84)90237-5}}.

\bibitem{Lee:2010wea}
R.~N. Lee, {Calculating multiloop integrals using dimensional recurrence
  relation and $D$-analyticity}, Nucl. Phys. Proc. Suppl. 205-206 (2010)
  135--140.
\newblock \href {http://arxiv.org/abs/1007.2256} {\path{arXiv:1007.2256}},
  \href {http://dx.doi.org/10.1016/j.nuclphysbps.2010.08.032}
  {\path{doi:10.1016/j.nuclphysbps.2010.08.032}}.

\bibitem{MR1633290}
B.~Bollob{\'a}s, \href{http://dx.doi.org/10.1007/978-1-4612-0619-4}{Modern
  graph theory}, Vol. 184 of Graduate Texts in Mathematics, Springer-Verlag,
  New York, 1998.
\newblock \href {http://dx.doi.org/10.1007/978-1-4612-0619-4}
  {\path{doi:10.1007/978-1-4612-0619-4}}.
\newline\urlprefix\url{http://dx.doi.org/10.1007/978-1-4612-0619-4}

\bibitem{Mastrolia:2016dhn}
P.~Mastrolia, T.~Peraro, A.~Primo, {Adaptive Integrand Decomposition in
  parallel and orthogonal space}, JHEP 08 (2016) 164.
\newblock \href {http://arxiv.org/abs/1605.03157} {\path{arXiv:1605.03157}},
  \href {http://dx.doi.org/10.1007/JHEP08(2016)164}
  {\path{doi:10.1007/JHEP08(2016)164}}.

\bibitem{Lee:2014tja}
R.~N. Lee,
  \href{https://inspirehep.net/record/1297497/files/arXiv:1405.5616.pdf}{{Modern
  techniques of multiloop calculations}}, in: {Proceedings, 49th Rencontres de
  Moriond on QCD and High Energy Interactions: La Thuile, Italy, March 22-29,
  2014}, 2014, pp. 297--300.
\newblock \href {http://arxiv.org/abs/1405.5616} {\path{arXiv:1405.5616}}.
\newline\urlprefix\url{https://inspirehep.net/record/1297497/files/arXiv:1405.5616.pdf}

\bibitem{Kosower:2011ty}
D.~A. Kosower, K.~J. Larsen, {Maximal Unitarity at Two Loops}, Phys. Rev. D85
  (2012) 045017.
\newblock \href {http://arxiv.org/abs/1108.1180} {\path{arXiv:1108.1180}},
  \href {http://dx.doi.org/10.1103/PhysRevD.85.045017}
  {\path{doi:10.1103/PhysRevD.85.045017}}.

\bibitem{Johansson:2012sf}
H.~Johansson, D.~A. Kosower, K.~J. Larsen, {An Overview of Maximal Unitarity at
  Two Loops}[PoSLL2012,066(2012)].
\newblock \href {http://arxiv.org/abs/1212.2132} {\path{arXiv:1212.2132}}.

\bibitem{CaronHuot:2012ab}
S.~Caron-Huot, K.~J. Larsen, {Uniqueness of two-loop master contours}, JHEP
  1210 (2012) 026.
\newblock \href {http://arxiv.org/abs/1205.0801} {\path{arXiv:1205.0801}},
  \href {http://dx.doi.org/10.1007/JHEP10(2012)026}
  {\path{doi:10.1007/JHEP10(2012)026}}.

\bibitem{Johansson:2012zv}
H.~Johansson, D.~A. Kosower, K.~J. Larsen, {Two-Loop Maximal Unitarity with
  External Masses}, Phys.Rev. D87 (2013) 025030.
\newblock \href {http://arxiv.org/abs/1208.1754} {\path{arXiv:1208.1754}},
  \href {http://dx.doi.org/10.1103/PhysRevD.87.025030}
  {\path{doi:10.1103/PhysRevD.87.025030}}.

\bibitem{Johansson:2013sda}
H.~Johansson, D.~A. Kosower, K.~J. Larsen, {Maximal Unitarity for the Four-Mass
  Double Box}\href {http://arxiv.org/abs/1308.4632} {\path{arXiv:1308.4632}}.

\bibitem{Sogaard:2013fpa}
M.~Sogaard, Y.~Zhang, {Multivariate Residues and Maximal Unitarity}, JHEP 12
  (2013) 008.
\newblock \href {http://arxiv.org/abs/1310.6006} {\path{arXiv:1310.6006}},
  \href {http://dx.doi.org/10.1007/JHEP12(2013)008}
  {\path{doi:10.1007/JHEP12(2013)008}}.

\bibitem{Hauser1993}
H.~Hauser, G.~M{\"u}ller, \href{http://dx.doi.org/10.1007/BF03026556}{Affine
  varieties and lie algebras of vector fields}, manuscripta mathematica 80~(1)
  (1993) 309--337.
\newblock \href {http://dx.doi.org/10.1007/BF03026556}
  {\path{doi:10.1007/BF03026556}}.
\newline\urlprefix\url{http://dx.doi.org/10.1007/BF03026556}

\bibitem{opac-b1094391}
D.~A. Cox, J.~B. Little, D.~O'Shea,
  \href{http://opac.inria.fr/record=b1094391}{Using algebraic geometry},
  Graduate texts in mathematics, Springer, New York, 1998.
\newline\urlprefix\url{http://opac.inria.fr/record=b1094391}

\bibitem{DiVita:2014pza}
S.~Di~Vita, P.~Mastrolia, U.~Schubert, V.~Yundin, {Three-loop master integrals
  for ladder-box diagrams with one massive leg}, JHEP 09 (2014) 148.
\newblock \href {http://arxiv.org/abs/1408.3107} {\path{arXiv:1408.3107}},
  \href {http://dx.doi.org/10.1007/JHEP09(2014)148}
  {\path{doi:10.1007/JHEP09(2014)148}}.

\bibitem{Faugere:2002:NEA:780506.780516}
J.~C. Faug\`{e}re, \href{http://doi.acm.org/10.1145/780506.780516}{A new
  efficient algorithm for computing gr\"{o}bner bases without reduction to zero
  (f5)}, in: Proceedings of the 2002 International Symposium on Symbolic and
  Algebraic Computation, ISSAC '02, ACM, New York, NY, USA, 2002, pp. 75--83.
\newblock \href {http://dx.doi.org/10.1145/780506.780516}
  {\path{doi:10.1145/780506.780516}}.
\newline\urlprefix\url{http://doi.acm.org/10.1145/780506.780516}

\end{thebibliography}







\end{document}